\documentclass[12pt]{article}
\usepackage{epsfig}
\usepackage{epsf,afterpage}
\usepackage{amsfonts,amssymb}

 \def\eeq{\end{eqnarray}}
 \def\beqs{\begin{eqnarray*}}
 \def\eeqs{\end{eqnarray*}}
 \def\dl{\delta}
 \def\la{\mathrel{\mathpalette\fun <}}

\def\fun#1#2{\lower3.6pt\vbox{\baselineskip0pt\lineskip.9pt
\ialign{$\mathsurround=0pt#1\hfil
##\hfil$\crcr#2\crcr\sim\crcr}}}
\newcommand{\be}{\begin{equation}}
 \newcommand{\ee}{\end{equation}}
 \newcommand{\lan}{\langle}
 \newcommand{\rrr}{\rangle}

 \newcommand{\T}{\mbox{Tr}\> }
\newcommand{\dv}{\mathrm{div}}
\newcommand{\rot}{\mathrm{rot}}

\newcommand{\veR}{\mbox{\boldmath${\rm R}$}}

\newcommand{\veB}{\mbox{\boldmath${\rm B}$}}
\newcommand{\veE}{\mbox{\boldmath${\rm  E}$}}
\newcommand{\vecE}{\mbox{\boldmath${\rm{\cal E}}$}}
\newcommand{\vecB}{\mbox{\boldmath${\rm{\cal B}}$}}
\newcommand{\ven}{\mbox{\boldmath${\rm n}$}}

\newcommand{\ver}{\mbox{\boldmath${\rm r}$}}
\newcommand{\vex}{\mbox{\boldmath${\rm x}$}}
\newcommand{\vek}{\mbox{\boldmath${\rm k}$}}

\newcommand{\vnabla}{\mbox{\boldmath${\rm \nabla}$}}
\newcommand{\bc}{\begin{center}}
\newcommand{\ec}{\end{center}}

\newcommand{\mr}[1]{\mathrm{#1}}
\newcommand{\fr}[2]{\frac{#1}{#2}}
\newcommand{\lt}{\left}
\newcommand{\rt}{\right}

\newcommand{\rf}[1]{(\ref{#1})}
\newcommand{\lb}{\label}

 \def\centeron#1#2{{\setbox0=\hbox{#1}\setbox1=\hbox{#2}\ifdim
 \wd1>\wd0\kern.5\wd1\kern-.5\wd0\fi
 \copy0\kern-.5\wd0\kern-.5\wd1\copy1\ifdim\wd0>\wd1
 \kern.5\wd0\kern-.5\wd1\fi}}
 \def\ltap{\;\centeron{\raise.35ex\hbox{$<$}}{\lower.65ex\hbox{$\sim$}}\;}
 \def\gtap{\;\centeron{\raise.35ex\hbox{$>$}}{\lower.65ex\hbox{$\sim$}}\;}

\begin{document}

 \vskip 1.2cm

\begin{titlepage}

 \vskip 1.2cm

 \begin{center}

 {\LARGE\bf The QCD vacuum, confinement and strings}

\vskip 0.1cm

{\LARGE\bf  in the Vacuum Correlator Method}
 \vskip 1.4cm
 {\bf \large D.S.Kuzmenko, V.I.Shevchenko, Yu.A.Simonov}
 \\
 \vskip 0.3cm
{\it Institute of Theoretical and Experimental Physics,
 \\ Moscow, Russia}
 \\
 \vskip 0.45cm 
 e-mail: kuzmenko@heron.itep.ru; shevchen@heron.itep.ru; simonov@heron.itep.ru

 \vskip 2cm

 \begin{abstract}
In this review paper the QCD vacuum properties and the structure of
color fields in hadrons
are studied using complete set of gauge-invariant correlators of
the gluon fields. Confinement in QCD is produced by the correlators of
some certain Lorentz structure, which violate abelian Bianchi identities
and therefore are absent in the case of QED. These correlators are used
to define an effective colorless field, which satisfies Maxwell equation
with nonzero effective magnetic current. With the help of the effective field
and correlators it is demonstrated that quarks are confined due to 
effective magnetic currents, squeezing gluonic fields into a string, in 
agreement with the ``dual Meissner effect''. Distribution of 
effective gluonic fields are
plotted in mesons, baryons and glueballs with static sources.
\end{abstract}  
\end{center} 
 \vskip 1.0 cm

 \end{titlepage}

 \setcounter{footnote}{0} \setcounter{page}{2}
 \setcounter{section}{0} \setcounter{subsection}{0}
 \setcounter{subsubsection}{0}

 
\section{Introduction}
The QCD is a unique example of field theory, lacking internal contradictions 
and at the same time explaining all physical phenomena in strong interactions 
 \cite{i1,i2}. The theoretical understanding of QCD is complicated due to the 
fact that all its basic features are of the nonperturbative nature, and the QCD
vacuum is a dense and highly nontrivial substance. In fact, in the modern
quantum field theory one often represents the vacuum as a specific material
substance with definite characteristics in direct analogy with the condensed 
matter physics. As illustrating examples one can mention the Casimir effect
and relative phenomena, and also the Higgs mechanism in the standard model.
In the last case one deals with the vacuum condensate of the scalar field
 $\lan \phi \rrr$, while quantum  excitations above this condensate are
considered as Higgs particles. 

The nontriviality of the QCD vacuum is revealed by the fact that this medium
has nonzero values of gluonic condensate \cite{i3}, $\lan F^a_{\mu\nu} F^a_{\mu\nu} \rrr = 
(600\>{\mbox{MeV}})^4$, and of the quark condensate,  $\lan \bar{q} q 
\rrr = - (250\>{\mbox{MeV}})^3$. As it has become clear during last decades
it is the vacuum properties which bring about confinement (see, e.g., the
review \cite{i4}). For theoretical calculations in QCD one usually exploited
till recently the perturbation theory augmented by some models of nonperturbative
mechanisms. The situation changed with the advent of the QCD sum rule method
 \cite{i3}, which uses the gauge-invariant formalism of condensates to describe
the nonperturbative contributions. However for most effects at large distances 
this method is not sufficient, e.g. for confinement or spontaneous violation
of chiral symmetry. The systematic  description of all, in principle,
QCD phenomena is made possible due to the appearence of the Vacuum Correlator
Method (VCM), see  \cite{i51,i52,i53} and the review \cite{i6}, which exploits
as basic elements the complete set of field correlators of the form 
\be
D^{(n)}_{\mu_1\nu_1 ... \mu_n\nu_n}(x_1,...,x_n,x_0) =
\lan \T G_{\mu_1\nu_1}(x_1,x_0) . . . G_{\mu_n\nu_n}(x_n,x_0) \rrr
\label{ir5}
\ee
where the notation   $G_{\mu_1\nu_1}(x_1,x_0)$ is used for the gluonic field
strength covariantly shifted along some curve, see Eq.   (\ref{ev983}).

The basis of the VCM are the gauge-invariant Green's functions of white objects,
which can be written as path integrals through field correlators (\ref{ir5})
using the cluster expansion, see e.g. \cite{i53,i71,i72}. A question may 
arise at this point, why one considers in VCM only white (i.e. 
gauge-invariant)
objects, and not for example propagators in some fixed gauge? The answer
is tightly connected to the difference between gauge invariance in abelian
and nonabelian theories. In the abelian theory, e.g. in QED, the
requirement of gauge-invariance
does not forbid to consider the problems with formally 
gauge-noninvariant asymptotic states, like electron-electron scattering.
The gauge invariance of the cross-section occurs in this case due to the
conservation of the abelian current. In the nonabelian theory with confinement,
like QCD, the situation is different and the problem of scattering of isolated 
quarks has no sense. Formally one can see that nonlocal gauge noninvariant matrix element
vanishes when being averaged over gluonic vacuum, which is a property of the
nonabelian $SU(N)$ group. Therefore instead one considers in QCD the quark-quark
 scattering for quarks inside white objects (i.e. described by the gauge-invariant
functions), such as hadrons. The same is true for the problems connected
to the spectrum of bound states --- while in QED the problem of a neutral
atom spectrum is just as valid as that of the spectrum of a charged ion, in QCD
an analog of the last problem has no meaning. 

Therefore the set of correlators  (\ref{ir5}) can be considered as a starting
dynamical basis yielding a phenomenological gauge-invariant description
of physical processes. In fact, however, the situation is much more interesting.
First of all, the lattice calculations give important evidence that already
the first nontrivial correlator with $n=2$ dominates, and the total contribution
of all highest correlators is below few percent, see review \cite{i9}. As was
shown in \cite{i52,i53}, the lowest (or Gaussian\footnote{using the analogy with
so-called Gaussian, or white, noise described by a quadratic correlator,
in this case with vanishing correlation length} as it will be called in what 
follows) correlator can be expressed through two scalar formfactors
 $D(x_1 - x_2)$ and $D_1(x_1 - x_2)$.   
Secondly, both formfactors  have been measured in the lattice calculations
and have nonperturbative parts of exponential shape with a characteristic
small correlation length $\lambda$. Finally, the function
 $D^{(2)}$ (and therefore also  $D(x_1 - x_2)$, $D_1(x_1 - x_2)$) are
directly connected to the Green's functions of the so-called gluelumps
 \cite{i111,i112,i113}. The latter can be calculated analytically in terms
of the only mass scale of QCD, e.g. through the string tension  $\sigma$ and
the coupling constant $\alpha_s$. Thus the formulation of the nonperturbative
dynamics in QCD turns out to be selfconsistent and one should in addition
calculate $\lambda$ through $\sigma$, what was done earlier in  \cite{i12},
and also connect  $\sigma$ and $\Lambda_{QCD}$ and write down the explicit form
of correlators  $D^{(n)}$. This way one would also be able to
understand analytically the dominance of $D^{(2)}$ (one can find the
first results in this direction in \cite{i13,i14}).

With all that the formalism of field correlators is to a large extent unusual
to physicists, brought up in the standard lore of perturbative, or even more,
of abelian gauge theory. In the context of the confinement problem such
a ``linear'' abelian approach is realized in the so-called ``dual Meissner 
scenario'', which contains a simple qualitative picture of the confinement 
mechanism in QCD \cite{i151,i152}. In this approach the acting roles have charges
(quarks) and the monopole medium filling the vacuum. Many lattice and analytic
studies, see, e.g.,  \cite{i16} - \cite{i182}, demonstrate that the string 
formation between quark and antiquark is connected in this picture with the
appearence of circular monopole currents  ${\bf k}$ around the string, which
obey the dual Ampere law ${\bf k} = {\mbox{rot}}\> {\bf E}$. From the physical
point of view this situation is similar to the Meissner effect in the standard
superconductivity phenomenon, modulo interchange of effective electric and
magnetic charges. On the other hand, the defect of this picture is that the very
notion of the magnetic monopole cannot be exactly defined in QCD. This 
arbitrariness  can be seen, first of all, in the gauge dependence of the
monopole definition, and secondly,
in the difficulties with the continuum limit for the lattice monopoles,
defined by the flux through an elementary cube. There is a lot of literature,
with different suggestions on how to deal with this problems, see e.g. \cite{i19}.

While confinement properties are studied on the lattice numerically, including
partly the abelian projection, they are also an object of investigation
in the effective Lagrangian approach and in different dielectric vacuum models
of QCD \cite{ap}-\cite{chrom2}. The basic field theory problem in this case
is replaced by a classical variational problem for the effective Lagrangian,
which yields a system of differential equations, to be solved numerically.
In this way one introduces an effective dielectric constant of the vacuum,
depending on the effective fields and ensuring quark confinement.

In what follows we shall use another approach which is fully 
gauge-invariant and yields a simple and selfconsistent picture of the 
confining string formation. Namely, using the field correlator method as a 
universal 
language, one can define
gauge-invariant (with respect to the gauge symmetry of the original nonabelian
theory) effective field $\:$ ${\cal F}_{\mu\nu}(x)$ via the W-loop.
The effective electric field near the charge turns out to be the gradient
of the color-Coulomb field, and in the case of an abelian theory
 ${\cal F}_{\mu\nu}(x)$ is the standard field strength. The effective field
satisfies Maxwell equations, having on the r.h.s. electric current
 $j_{\mu}$ and magnetic current $k_{\mu}$. The source of $k_{\mu}$
is primarily the triple correlator of the form $\lan EEB \rrr$ (as was already
 found in \cite{i4}) describing the emission of the color-magnetic field by the 
color-electric; the latter can be visualized as the emission of the color-magnetic
field by an effective magnetic charge (monopole). In the language of field 
correlators one can easily demonstrate that the system of equations for the 
effective fields describes the QCD string and the circular magnetic currents around
 it. In this way the picture of the dual Meissner effect is given in 
gauge-invariant terms.

With the help of  ${\cal F}_{\mu\nu}$ one can investigate in detail the 
structure
of the QCD string. The first computations of the string profile in  \cite{i20}
have demonstrated a very good agreement of the results calculated via $D(x^2), 
D_1(x^2)$ and those obtained independently on the lattice. The following
study of the string structure \cite{field} has shown an interesting phenomenon
of the profile saturation, where the profile (i.e. the field distribution
across the string) does not change for long enough strings. The relief of the
baryon field has turned out to be even more interesting. Baryons,
and more exact, nucleons are the basis of the bulk of the stable matter
around us. The physical problem of the structure of the baryon field
is especially interesting both
from theoretical and practical points of view.
Two types of baryon field configurations were discussed in the literature: with the string junction in the middle
(the $Y$-shape) and of the triangular shape (the ($\Delta$-shape).
Using vacuum correlator method the baryon configuration was computed analytically
in \cite{string,rep3}, where the presence  of the string junction in the field
distribution was explicitly  demonstrated, thereby excluding the $\Delta$-type
configuration. On the other hand, the latter is possible for the three-gluon
glueballs and the corresponding field was calculated in  \cite{rep3}. One should
mention that these baryon field distributions are also in agreement with
the lattice calculations using the abelian projected QCD \cite{Born1}, see also
the review paper of Bornyakov {\it et. al.} \cite{Born2}.

The field sources in three-gluon glueballs are three valence gluons.
The field structure of these systems has some specific features, and
can be of both types, of the $\Delta$-type (unlike baryons), and of the
$Y$-type  (like baryons), and its study helps to understand better physics
of confinement. Moreover, the three-gluon glueballs have to do with the
processes of the odderon exchange (i.e. glueball exchange with odd charge parity),
and hence are also interesting from the experimental point of view.
Therefore, in addition to the effective field distributions, we shall also 
discuss below the W-loops and the static potentials of baryons and three-gluon
 glueballs.  

The paper has the following structure. In chapter 2 the discussion of field
correlator properties in QCD is given, and in particular the important
phenomenon of the Casimir scaling is explained. In chapter 3 the effective
field ${\cal F}_{\mu\nu}$ and currents $j_{\mu}$, $k_{\mu}$ are introduced
and the dual Meissner effect is demonstrated. In chapter 4 the static potentials
and field  distributions in baryons and three-gluon glueballs are given.
In Conclusions the main results are summarized and some prospectives are outlined.

Everywhere in what follows, if it is not especially stressed otherwise,
the Euclidean metrics is used with notations for 4-vectors 
$k=(k_1,k_2,k_3,k_4)$
and for scalar products $kp = k_{\mu} p^{\nu} \delta_{\nu}^{\mu}$.
The three-dimensional vectors are denoted as  ${\bf k} = (k_1,k_2,k_3)$ and
the Wick rotation corresponds to the replacement $k_4 \to ik_0$.

 \section{Properties of QCD vacuum in gauge-invariant approach}

\subsection{Definition of gauge-invariant correlators}

The following remark is to be made before we proceed. There is an
important difference between pure Yang-Mills theory (gluodynamics)
and QCD, namely the latter contains dynamical fermions, in
particular light $u$ and $d$ quarks. This circumstance plays no
crucial role in the description of confinement since gluodynamics
confines color as QCD does, which is supported by direct lattice
calculations (see, e.g. \cite{bali1}) and different qualitative
arguments. Because of that in most cases we consider pure
Yang-Mills theory in this review, while quarks play a role of
external sources.

\noindent One of the main objects in gauge theory is the Wegner-Wilson
loop \cite{w1,w2} which we denote here as W-loop: 
\be W(C)= {\mbox{P}}\exp i
g\oint\limits_C A_{\mu}^a(z) t^a dz_{\mu} \label{ev981} \ee where
$t^a$ - generators in the given representations of the gauge
group. W-loop defines external current $J$ which corresponds
to a point particle charged according to the chosen representation
and moving along the closed contour $C$. Phase factor for
non-closed curve connecting points $x$ and $y$ is also of
importance \be \Phi(x;y)= {\mbox{P}}\exp i g\int\limits_x^y
A_{\mu}^a(z) t^a dz_{\mu} \label{ev982} \ee Under the gauge
transformations we have \be \Phi(x;y) \to \Phi^U(x;y) =
U^{\dagger}(x) \Phi(x;y) U(y) \label{err} \ee It means that the
trace $\T W(C)$ is gauge-invariant.\footnote{In the literature the
trace is often included in the definition of the W-loop.} We
normalize $\T$ everywhere as $\T {\bf 1}_d = 1$ for the given
representation of dimension $d$. Making use of the  definition
(\ref{ev982}), let us introduce  $G_{\mu\nu}(x,x_0)$ as \be
G_{\mu\nu}(x,x_0) = \Phi(x_0;x)F_{\mu\nu}(x)\Phi(x;x_0)
\label{ev983} \ee where $F_{\mu\nu} = \partial_{\mu} A_{\nu} -
\partial_{\nu} A_{\mu} -i[A_{\mu}A_{\nu}]$ is nonabelian field
strength and the curve connecting the points $x$ and $x_0$ does
not self-intersect. In abelian theory $G_{\mu\nu}(x,x_0)\equiv
F_{\mu\nu}(x)$, however in Yang-Mills theory $G_{\mu\nu}(x,x_0)$
and $F_{\mu\nu}(x)$ transform differently under gauge
transformations, as it is clear from (\ref{err}). We can now
construct vacuum averages of the products of $G_{\mu\nu}(x_n,x_0)$
in the following way \be D^{(2)}_{\mu\nu\rho\sigma}(x,y,x_0) =
\lan \T G_{\mu\nu}(x,x_0) G_{\rho\sigma}(y,x_0) \rrr
\label{ev9884} \ee \be
D^{(3)}_{\mu\nu\rho\sigma\alpha\beta}(x,y,z,x_0) = \lan \T
G_{\mu\nu}(x,x_0) G_{\rho\sigma}(y,x_0) G_{\alpha\beta}(z,x_0)
\rrr \label{ev9885} \ee and analogously for higher orders. The
correlators (\ref{ev9884}), (\ref{ev9885}) are gauge-invariant,
but nonlocal - expressions (\ref{ev9884}), (\ref{ev9885}) depend
on the position of the points $x$, $y$, $z$ as well as on the
position of the point $x_0$ and contour profile used in
(\ref{ev983}). Physical observables such as static potential
extracted from the W-loop do not depend on $x_0$ and contour
profiles when all correlators $D^{(n)}$, $n\ge 2$ are taken into
account. It is not true, however if one takes only the lowest
$n=2$ term. In this case it is convenient to minimize the
corresponding dependence like one does in perturbation theory
minimizing the contribution of omitted terms by the proper choice
of  subtraction point $\mu$ on which the exact answer should not
depend.

\subsection{Computation of the W-loop and Green's functions in terms of
correlators}

Speaking in general terms, for a given gauge theory each function $D^{(n)}$
is important characteristics of its vacuum structure by itself. What is
more important, however is the possibility to express W-loop average in 
terms of correlators (\ref{ev9884}), (\ref{ev9885}).
Indeed, Stokes theorem (or, more precisely, its nonabelian generalization \cite{nast1}-\cite{nast6}) leads to \be
\lan \T W(C) \rrr = \left\lan \T {\cal P}\exp ig\int\limits_S G_{\mu\nu}(z,x_0) d\sigma_{\mu\nu}(z) \right\rrr
= \exp \sum\limits_{n=2}^{\infty} (ig)^n \Delta^{(n)}[S]
\label{nast}
\ee
there we have used cluser expansion to exponentiate the series (see, e.g.
\cite{vk1,vk2}). Integral moments $\Delta^{(n)}[S]$ over the surface $S$ of {\it irreducible} correlators, known as cumulants in statistical physics can be expressed as linear combinations of the integrals of correlators
 $D^{(n)}$. For example, we have for two-point correlator
\be
\Delta^{(2)}[S] = \frac12 \int\limits_S d\sigma_{\mu\nu}(z_1)
\int\limits_S d\sigma_{\rho\sigma}(z_2) D^{(2)}_{\mu\nu\rho\sigma}(z_1,z_2,x_0)
\label{ev960}
\ee
For higher terms the ordering is important, see, e.g. \cite{i13},
where exact computations for $n=4$ are performed.

Expression (\ref{nast}) is of central importance for the discussed
formalism. Let us consider the propagation of the spinless
particle with mass $m$, carrying fundamental color charge
("quark") in the field of infinitely heavy "antiquark" \cite{i53,i71,i72}.
The corresponding gauge-invariant Green's function reads as \be
{\cal G}(x,y) = \lan \phi^{\dagger}(x) \Phi(x;y) \phi(y) \rrr
\label{ev156} \ee where we denote quark field as $\phi(x)$. One
can demonstrate that ${\cal G}(x,y)$ has the following
Feynman-Schwinger representation \be {\cal G}(x,y) =
\int\limits_0^{\infty} ds \int\limits_{z_\mu(0) = x_\mu}^{z_\mu(s)
= y_\mu} {\cal D} z_{\mu} \exp\left(-m^2s -\frac14 \int\limits_0^s
d\tau \left(\frac{dz_{\mu}(\tau)}{d\tau}\right)^2 \right) \cdot
\lan \T W(C) \rrr \label{ev157} \ee where the closed contour $C$
is formed by the quark trajectory $ z_{\mu}(\tau)$ and that of
antiquark (the latter is nothing but the straight line connecting
the points  $x$ and $y$). We have taken spinless case here as the
simplest illustrative example, for real physical problems with
spinor quark fields there is a systematic way of analysis of spin
effects \cite{simlq,i71,i72}. The problem of two-body meson state or
three-body baryon one can be addressed in completely analogous
way. In all cases Green's function containing full information
about mass spectrum and wave functions of the system can be
re-written in terms of path integrals of the W-loops there
the latter are expressed via correlators  as in (\ref{nast}).

Therefore the set of correlators
 $D^{(n)}$  provides rich and, what is more important, universal dynamical information one can use to compute
 different nonperturbative effects.\footnote{Discussed formalism can be
 applied in perturbation theory as well. In this context it allows to sum up perturbative
 subseries with subsequent exponentiation with the well-known "Sudakov formfactor" as a result of the first
 approximation, see \cite{yu8}, \cite{i72} and references therein.}
Let us stress once again that the correlator (\ref{ev9884}) is
itself related to the Green's function of gluon excitation in the
field of infinitely heavy adjoint source - known in the literature
as gluelump \cite{i111}- \cite{i113}.

Coming to practical side of the problem, it is natural to ask what
the actual behavior of the correlators  (\ref{ev9884}),
(\ref{ev9885}) is and how information about it can be gained. This
question is simple to answer in perturbation theory since each
$D^{(n)}$ is given by perturbative series, see, e.g. \cite{ej1,ej2}.
There are a few ways to proceed beyond perturbation theory. The
first one is to find nonperturbative solutions to the so called
BBGKI equations, relating the correlators of different orders
\cite{as}. This way has brought no essential progress up to now.
Another analytic strategy suggests to compute correlators in terms
of gluelump Green's functions \cite{i111}-\cite{i12}. The third and the
most successful way is to study the problem on the lattice. There
are quite a few sets of numerical data \cite{i101}-\cite{bcor}, which we
discuss below. However it is obvious that numerical results
concerning one or a few particular correlators are useless if
general properties of the whole ensemble are unknown. To discuss
them we come back to the expression (\ref{nast}).

\subsection{Gaussian dominance}

It has already been stressed that the price we have payed for
manifest gauge-invariance of (\ref{nast}) is the dependence of
(\ref{ev9884}), (\ref{ev9885}) on the contour profiles entering
$\Phi(x;y)$. These contours are, generally speaking, arbitrary
non-selfintersecting curves or, better to say, they can be freely
chosen in some (large enough) set. As a result the quantities
$\Delta^{(n)}[S]$ in (\ref{nast}) depend on this choice while
$W(C)$ is obviously independent on $S$. The contradiction is
spurious and one can demonstrate that this contour dependence is
cancelled in the total sum, despite it is present in each
individual summand $\Delta^{(n)}[S]$. In this sense the choice of
the surface $S$ in (\ref{nast}) (corresponding to the choice of
integration contours in the correlators $D^{(n)}$) is free, as it
should be. We can take a different attitude and ask the following
question: what is the hierarchy of cumulants $\Delta^{(n)}[S]$ on
some particular surface? This question is of general interest but
it has also important practical meaning - in many problems one has
to deal with the surface, which is singled out by some physical
reasons. For a single W-loop it is obviously given by minimal
area surface, bounded by the contour. In more complicated case of
interacting loops  \cite{we5} the surface corresponding to the
minimal energy of the system can be taken. In any case, it is
instructive to make a distinction between two different scenarios:
\be
 \Delta^{(2)}[S] \gg \sum\limits_{n=3}^{\infty}\>\Delta^{(n)}[S]
\label{ev65} \ee which is referred to as {\it stochastic}
scenario, while the case when (\ref{ev65}) does not hold (for
example, all cumulants are of the same order) is known as {\it
coherent}. General framework described in the present paper takes
into account effects of all cumulants but as it should be clear,
it shows its strong sides in stochastic case. The lowest two-point
Gaussian cumulant (\ref{ev960}) is dominant in stochastic
ensemble, while higher order terms can be considered as small
corrections. This situation is known as Gaussian dominance. Then
we can ask is the QCD vacuum stochastic or coherent? To answer
this question in straightforward way one has to compute (for
example, numerically on the lattice) different cumulants and check
them against (\ref{ev65}). Unfortunately this research program is
too intricate for modern lattice technologies and almost all
actual results are obtained for Gaussian cumulant only. There are
important indirect evidences however supporting the idea that Yang
-Mills vacuum is indeed stochastic and not coherent in the sense
of (\ref{ev65}). Of prime importance in this context is Casimir
scaling phenomenon
 \cite{bali21}-\cite{deldar2}, see also \cite{lucini1}-\cite{lucini4}. Using (\ref{nast})
and taking into account well known relation between static
potential and W-loop average, one can get, assuming Gaussian
dominance \be V(R)= \lim\limits_{T\to\infty} \frac{1}{T} \> g^2
\Delta^{(2)}[S=R\times T] \label{ev36} \ee and, according to
(\ref{ev9884}) and (\ref{ev960}) we have $V(R)\sim C_d$, where an
eigenvalue of Casimir operator in the representation $d$ is given
by  $\dl_{ab} t^a t^b  = C_d \cdot {\bf 1}_d $. Let us remind that
representation of the Lie group $SU(N)$ of dimension $d$ is
characterized by $N^2 -1$ generators  $t^a$, which can be realized
as $d\times d$ matrices commuting as $[t^a t^b] = if^{abc} t^c$.
Proportionality of the static potential to $C_d$ is called Casimir
scaling \cite{faber3} and was discussed for the first time in \cite{aop}.

It can easily be shown that contributions from higher cumulants to the static potential (\ref{ev36}) are, generally speaking, not proportional to
 $C_d$ (despite they can contain linear in $C_d$ terms). Therefore a good accuracy (deviation not exceeding 5\%) of Casimir scaling demonstrated on the lattice
is a serious argument in favor of Gaussian dominance. Moreover,
attempts to reproduce Casimir scaling in many other models of
nonperturbative QCD vacuum encounter difficulties
\cite{simcas11,simcas12,i9}. Another argument is the observed independence
of radius of the confining string between quarks on their
nonabelian charge (i.e. on representation $d$) \cite{trot}. These
results would look as fine tuning effects without Gaussian
dominance. It is also worth mentioning that "vacuum state
dominance" successfully used for years in QCD sum rules formalism is nothing
but Gaussian dominance in our language.

\subsection{Structure of two-point correlators}

We have mentioned above the relation between correlators
 $D^{(n)}$ and gluelump Green's functions. For the simplest Gaussian correlator (\ref{ev9884}) this can be seen clearly if the contours are straight lines and points $x,y,x_0$ belong to one and the same line. The correlator depends on the only variable $z=x-y$ in this case and can be represented as
\be
D^{(2)}_{\mu\nu\rho\sigma}(z) =
\left\lan F_{\mu\nu}^a(0) \cdot {\mbox{P}}\exp \left(ig\int\limits_0^1 ds\>
z_{\mu} A^b_{\mu}(sz)f^{abc}\right)\cdot F^c_{\rho\sigma}(z) \right\rrr
\label{ev9888}
\ee
Expression (\ref{ev9888}) contains phase factor in the adjoint 
(compare with the previous formulas where we worked with fundamental phase factors, i.e. with $N\times N$ matrices) which makes its physical content self-evident.
Namely, gauge-invariant function  $D^{(2)}(z)$ describes
gluon propagation in the field of infinitely heavy adjoint charge at the origin in full analogy with fundamental case (compare (\ref{ev156}) and (\ref{ev9888})).

Confining string worldsheet given by the surface $S$ in
(\ref{nast}) interacts with itself by gluelump exchanges. This
interaction depends on the profile of $S$ in such a way that the
total answer for the W-loop average is $S$-independent.
Gaussian dominance means, qualitatively, that for some particular
surface this "gluelump gas" becomes "ideal" and integral
contribution of higher cumulants $\Delta^{(n)}$, $n>2$  is small
on this surface. This also means that two-gluon gluelumps weakly
interact with each other. The deviation from Casimir scaling (as
we have already noticed, it is small) can be expressed in terms of
irreducible averages of gauge-invariant operators
 $\lan \T {\cal O}_1 \T {\cal O}_2 \rrr$, describing interaction of gluelumps \cite{i13}. One immediately realises that such deviation is suppressed in large $N$ limit. To avoid misunderstanding let us stress that gluelumps do not exist as physical particles in the spectrum of the theory. It would also be wrong to interpret  (\ref{ev9888}) in terms of "massive gluon". In a limited sense gluelumps are analogous to Kalb-Ramond fields which describe dual vector bosons and play important role in constructing string representation of compact QED \cite{polyakov} and abelian Higgs model \cite{lee}
 (see also \cite{ant2,ant3}).
The discussed picture with gluelump ensemble on the worldsheet makes sense only in the presence of external current, forming the W-loop. On the other hand the correlator (\ref{ev9888}) may be studied as it is, with no reference to any external source. Before we discuss actual lattice results, it is useful to represent (\ref{ev9888}) in terms of two invariant formfactors
 $D(z^2)$ and $D_1(z^2)$ \cite{i51}-\cite{i53}
$$
g^2 D^{(2)}_{\mu\nu\rho\sigma}(z) = (\dl_{\mu\rho} \dl_{\nu\sigma} -
\dl_{\mu\sigma} \dl_{\nu\rho}) D(z^2) +
$$
\be + \frac12 \left(\frac{\partial}{\partial z_{\mu}}  (z_{\rho}
\dl_{\nu\sigma} - z_{\sigma} \dl_{\nu\rho})  -
\frac{\partial}{\partial z_{\nu}}  (z_{\rho} \dl_{\mu\sigma} -
z_{\sigma} \dl_{\mu\rho})\right)  D_1(z^2) \label{ew1} \ee
Confinement (linear potential between static quark and antiquark
in fundamental representation) takes place in Gaussian dominance
picture when $D(z^2)$ is nonzero. At large distances we have from
(\ref{ev960}), (\ref{ev36}) for static potential $V(R)$ and string
tension $\sigma$: \be V(R) = \sigma R + {\cal O}(R^0) \;\;\; ;
\;\;\; \sigma = \frac12 \int d^2 z D(z^2) \label{potw2} \ee while
at small distances perturbative contribution dominates \cite{ej1,ej2}.
Nonperturbative part of the correlator is usually taken as \be
D(z^2) \sim \exp\left(-|z|/\lambda\right) \label{expon} \ee and this
exponential fit is in very good agreement with lattice data at
large enough distances. The situation with nonperturbative
component of the function $D_1(z^2)$ is less clear. In any case, 
physically the exact function $D_1(z^2)$ containing perturbative and
nonperturbative pieces must be exponentially
suppressed at large enough distances.
It is important that from practical point of view one has
no need to know the detailed profile of formfactors $D(z^2)$,
$D_1(z^2)$: physical quantities are given as integral moments of
these functions as in (\ref{potw2}). Quantity $\lambda$ is known as
correlation length of QCD vacuum and as it is clear from our
discussion this quantity is nothing but the inverse mass of the
lowest gluelump: $\lambda = 1/M$. On the other hand, typical size of
vacuum domain where fields are correlated is given by the same
$\lambda$ \cite{nacht}. We use numerical value $\lambda = 0.2$ fm in
accordance with the lattice results. Physics of nonlocality
switches on at distances larger that $\lambda$ and has many
phenomenological manifestations. One of the most interesting - the
confining string formation - will be discussed in what follows.

So far we have not mentioned the problem of deconfinement. There are basically two groups of physically interesting questions related to this problem. The first one covers dynamical aspects of the phase transition, while the second group deals with symmetric properties of the ground state (and excitations) in different phases. In the context of our discussion a typical question from the
first group looks like the following: what does temperature
deconfinement phase transition correspond to in terms of correlators?
The second group provides questions like: where is screening of zero
$N$-ality charges at large distances hidden in the expression
(\ref{nast})? We have no possibility to discuss these important issues
in the present review and refer the reader to original literature and
references therein  (see review \cite{i6}).

\section{Mechanism of confinement and dual Meissner effect}
\subsection{Effective fields definition}

The formalism considered so far allows one to perform the expansion of Wilson
loop  \rf{nast} and  static potential  \rf{ev36} over
the full set of field correlators \rf{ev9884}, \rf{ev9885}, \rf{ew1},
\rf{expon} in the whole range of distances. In what follows we will use
these results to calculate the effective confining field in hadrons and study
some of its phenomenological applications\footnote{Dynamics of effective
fields is considered in \cite{D1}, \cite{D2}}.

It is well-known that the static potential at small quark-antiquark distances 
$r\ll \Lambda_{QCD}$ in Born approximation of perturbation theory has the form
\be
V^{\mr{Coul}}(r)=-\fr{C_F \alpha_s}{r},
\lb{V1}
\ee
where $C_F=4/3$ is the quadratic Casimir operator in fundamental
representation. The color factor  $C_F$ is the only difference between this
potential and Coulomb one in electrodynamics. One can introduce the field
\be
\vecE^\mr{Coul}= \vnabla V^\mr{Coul}(r),
\lb{V2}
\ee
which has the meaning of the force acting on the quark. 

Let us define the effective field as follows,
\be
{\cal F}^J_{\mu\nu}(x) =
\lan \T W(C) \rrr^{-1} \lan \T ig G_{\mu\nu}(x,x_0) W(C)
\rrr.
\label{F}
\ee
Index $J$ stresses that the field ${\cal F}^J_{\mu\nu}(x)$ is the
functional of the external current $J$ corresponding to W-loop $W(C)$.
It will be demonstrated in the next section that this effective field at small
distances is reduced to color-Coulomb field  \rf{V1}, \rf{V2}.

Notice that one can write down the effective field using the {\it
  connected probe} \cite{dig5} $\lan \T W(C,C_P) \rrr$, where
\be
W(C,C_P)=W({C_P},x)\Phi(x,x_0)\Phi(x_0,z)
W(C,z)\Phi(z,x_0)\Phi(x_0,x)
\label{F2}
\ee 
is the W-loop with the contour consisting of the (small) probe
contour  $C_P$ connected with the contour $C$ along some trajectory
going through the point $x_0$.
This quantity depends on the position of the "reference point" $x_0$ as well as on the shape of 
the trajectory
connecting $C$ and $C_P$. We will choose the trajectory going along the shortest path from point $x$ to
the minimal surface of the W-loop, see Fig. 1.

The effective field in the case of probe contour $C_P$ with the infinitesimal
surface    $\delta \sigma_{\mu\nu}$  can be written as
\be
{\cal F}^J_{\mu\nu}(x)\,\delta \sigma_{\mu\nu}(x)=
\lan \T W(C) \rrr^{-1}\lt(\lan \T W(C,C_P) \rrr-
\lan \T W(C) \rrr\rt)\equiv \tilde M(C,C_P)
\label{F3}
\ee 
In particular, if the probe contour has a size $a\times a$, 
the relation for the electric field follows,
 \be
\ven\cdot \vecE^J(x)=\fr{\tilde M(C,C_P)}{a^2},
\label{F4}
\ee
where $\ven$ is the unit vector defining the orientation of probe contour
in coordinate space\footnote{Let us remind in this context the expression for the moment of
forces acting on the frame with the electric current $I$
in the magnetic field $\veB$, known from the general physics. Namely, 
when the frame is oriented in the
plane ($\ven^{(1)}$, $\ven^{(2)}$) and $\ven^{(1)}$ is chosen
orthogonal to magnetic field, the moment of acting forces $M$ takes the form
$
\ven^{(2)}\cdot \vecB=\fr{M}{a^2},
\label{F5}
$
where $\vecB\equiv I~\veB$. Comparing this relation with \rf{F5} one sees
that $\tilde M(C,C_P)$ defined in \rf{F3} means the "dual" moment of acting forces.}.

\subsection{Definition of effective currents}

In abelian theory
$G_{\mu\nu}(x,x_0)\equiv F_{\mu\nu}(x)$ and equation
\rf{F} defines field distribution satisfying  
Maxwell equations with the external electric
current $g^2 J_{\mu}(x) = g^2\int_C dz_{\mu}\delta^{(4)}(z-x)$
\be
\frac{\partial}{\partial x_{\rho}} {\cal F}^J_{\rho\mu}(x) = g^2 
J_{\mu}(x)
\label{uyi}
\ee
where $g$ denotes  the electric charge. Let us now proceed with the 
nonabelian case.
Using the differential relations for phase factors (see, e.g.,
 \cite{nast1} - \cite{nast6}), one can formally write down
 the effective "electric" and "magnetic" currents as
 $$
 j_{\nu}^J(x) =  \lan \T W(C) \rrr^{-1} \left\{ \vphantom{\int_0^1 ds \frac{\partial u_{\alpha}(s,x)}{\partial s}
\frac{\partial u_{\beta}(s,x)}{\partial x_{\mu}} }
\lan \T \Phi(x_0;x) ig D_{\mu}F_{\mu\nu}(x)\Phi(x;x_0) W(C) \rrr \right.
$$
\be
\left. + g^2 \int_0^1 ds \frac{\partial u_{\alpha}(s,x)}{\partial s}
\frac{\partial u_{\beta}(s,x)}{\partial x_{\mu}} \lan \T
[G_{\alpha\beta}(u,x_0) G_{\mu\nu}(x,x_0) ] W(C) \rrr \right\}
\label{ew92}
\ee
$$
k_{\nu}^J(x) = g^2 \lan \T W(C) \rrr^{-1}\times
$$
\be
 \int_0^1 ds 
\frac{\partial u_{\alpha}(s,x)}{\partial s} 
\frac{\partial u_{\beta}(s,x)}{\partial x_{\mu}} \lan 
\T  [G_{\alpha\beta}(u,x_0) {\tilde G}_{\mu\nu}(x,x_0) ]\> W(C) \rrr
\label{k1}
\ee   
where the integration contour  is given by the function $u_{\mu}(s,x)$ 
with the boundary conditions
 $u^\mu(0,x)=x_0^\mu$,  $u_\mu(1,x)=x_\mu$ and square brackets
denote commutators in color space.    
Index $J$ indicates that the "electric" current
 $j^J_{\mu}$ and "magnetic" one $k^J_{\mu}$ are functionals
 of the external current $J$ given by the W-loop. The currents 
defined in such a way 
can now be considered as sources of effective "electric" and "magnetic"
fields according to effective "Maxwell equations"
\be
\frac{1}{2}\>\epsilon_{\mu\rho\alpha\beta}
\frac{\partial}{\partial x_{\rho}} {\cal F}^J_{\alpha\beta}(x) =
k^J_{\mu}(x)\>\>\>\>; \>\>\>\>\>\>
\frac{\partial}{\partial x_{\rho}} {\cal
F}^J_{\rho\mu}(x) = j^J_{\mu}(x),
\label{Meq}
\ee

Equations (\ref{ew92}) and (\ref{k1}) define effective currents which
satisfy (\ref{Meq}) with the definition (\ref{F}) identically.
Notice that in (\ref{k1})
nonabelian Bianchi identities $D_{\mu}{\tilde F}_{\mu\nu}(x) =0$ 
respecting the gauge nature of QCD are used.
  It is obvious from  \rf{Meq} that both electric and magnetic
  effective currents are conserved 
since tensor ${\cal F}_{\mu\nu}$ is antisymmetric.

Let us note that the lowest term of the W-loop expansion,
which contributes    to    $k_{\mu}^J(x)$  (\ref{k1}), is proportional to
nonabelian field strength    correlator of third order.
Therefore the value of magnetic current is
proportional to correlator
$\lan E^a_iB^b_jE^c_k\rrr f^{abc}\epsilon_{ijk}$,
i.e. the effective magnetic current emerges due to the nonabelian
emittence of the colormagnetic field by the colorelectric one \cite{i4}.
The averages of the type $\lan  \T  G_{\alpha\beta}(x,x_0)
    G_{\gamma\delta}(y,x_0)\, W(C) \rrr$ in rhs of \rf{ew92}, \rf{k1}
define the nonlocal gluon condensate in the presence of W-loop,
which saturates to the constant value far from the W-loop.    
We do not address here an interesting question about possible microscopic 
nature of the currents (\ref{ew92}), (\ref{k1}), in particular, the question
to what extent the magnetic current (\ref{k1}) may be understood as
corresponding to some propagating point-like particles, "abelian monopoles".
Instead, we take  (\ref{ew92}), (\ref{k1}) as primary effective definitions.

If the gauge coupling is small, one can use
 for the electric current \rf{ew92}
the equation of classical gluodynamics,
\be
ig D_\mu F^a_{\mu\nu}=g^2\,J_\nu^a ,
\lb{j2}
\ee
where $J_{\mu}^a(x) =J_{\mu}(x)T^a$, ~
$J_{\mu}(x) =\int_C dz_{\mu} \delta^{(4)}(z-x)$.
In the leading order in gauge coupling $\alpha_s=g^2/(4\pi)$
the second term of \rf{ew92} does not contribute, and  the expression
for the electric current reads as
\be
j_{\nu}^J(x) = 4\pi C_F \alpha_s\, J_\nu(x),
\lb{j3}
\ee
i.e. it has a form of classical current of electrodynamics
with the charge $C_F \alpha_s$. In particular case
of static quark and antiquark Maxwell equation with this current reproduces the
color-Coulomb potential \rf{V1}.

\subsection{Effective fields distribution in two-point approximation}

Let us consider the rectangular W-loop of static quark and antiquark.
Relying on the  hypothesis of bilocal (gaussian) dominance  we take
into account of only the bilocal correlator
contribution to the effective fields
assuming that higher correlators do  not lead to
essential modification of the confinement picture. The effective field in
bilocal approximation reads 
\be
{\cal F}_{\mu\nu}(x)=\int_{S}
d\sigma_{\alpha\beta}(y)\, g^2 D^{(2)}_{\alpha\beta\mu\nu}(x-y),
\lb{3.4}
\ee
where $y\in S$,~ $S$ is the minimal surface of the W-loop, and bilocal
correlator $D^{(2)}$ is defined in \rf{ew1}.

Let us denote $\ven=\veR/R$ the unit vector directed from quark to antiquark
and rewrite \rf{3.4} in the form
\be
{\cal F}_{\mu\nu}(x)=\int_S d^2 y\,\T
\left\lan gF_{\mu\nu}(x)\Phi(x,y) \ven g\veE(y)\Phi(y,x) \right\rrr,
\lb{3.5}
\ee
which clearly indicates that the magnetic field ${\cal B}$ is absent.
The substitution of parametrization \rf{ew1} for \rf{3.5} yields the following
expression for the effective electric field,
\be
{\cal E}_i(\ver,\veR)=n_k\int\limits_0^R
dl\int\limits_{-\infty}^{\infty}dt \lt(\delta_{ik}D(z)+\fr12
\fr{\partial z_i D_1(z)}{\partial z_k} \rt),
\lb{Edist}
\ee
where $z=(\ver-\ven l,t)$.
The perturbative part of the field 
corresponding to the contribution of the formfactor $D_1$
 to  \rf{Edist} can be represented as the difference
\be
\vecE^{D_1,\mr{oge}}(\ver)=
\vecE^\mr{Coul}(\ver)-\vecE^\mr{Coul}(\ver-\veR),
\lb{3.8}
\ee
where $\vecE^\mr{Coul}(\ver)$ is the color-Coulomb field \rf{V1}, \rf{V2},
\be
\vecE^\mr{Coul}(\ver)=\fr{C_F\alpha_s \ver}{r^3}.
\lb{3.10}
\ee
The corresponding formfactor,
\be
D^\mr{oge}_1(z)=\fr{4C_F\alpha_s}{\pi z^4},
\lb{3.11}
\ee
can also be calculated directly in perturbation theory \cite{Shevchenko}.

It was discussed in previous chapter  that the confinement is the consequence
of stochastic nature of gluon field fluctuations, which reveal themselves
at separations of the order of the correlation length $\lambda$ and lead to the
exponential fall off of the field correlators, see \rf{expon}.
One can show that since at large separations the string acts on quark with
the force $\sigma$, the formfactor $D$ should be normalized according to
\be
D(z^2) = \fr{\sigma}{\pi \lambda^2} \exp\left(-\frac{|z|}{\lambda}\right).
\lb{3.12}
\ee
On substituting \rf{3.12} for \rf{Edist} one calculates the corresponding field,
\be
{\cal \vecE}^D(\ver,\veR)=\ven\, \fr{2\sigma}{\pi}
\int\limits_0^{R/\lambda} dl\, \lt| l\ven-\frac\ver{\lambda}\rt|
K_1\lt(\lt| l\ven-\frac\ver{\lambda}\rt|\rt),
\lb{3.13}
\ee
where $K_1$ is the McDonald function.
The string tension $\sigma$ can be considered as a scale QCD parameter
(it is related to $\Lambda_{\mr{QCD}}$ through equation \rf{3.27}).
Numerical value $\sigma\approx 0.18$ GeV$^2$ is determined phenomenologically
from the slope of the meson Regge trajectory, see e.g. \cite{hyb1}.
It is easy to verify that if the point $x$ is placed at the symmetry axes, the
relation between the field ${\cal \vecE}^D$ and the nonperturbative part of
the static potential corresponding to formfactor $D$ (\ref{ev960}),
\rf{ev36}, which we denote  $V^D$, reads
\be
\vecE^D(0,\veR)=\vnabla V^{D}(R).
\lb{3.13b}
\ee

The distribution of the field $|\vecE(x_1,0,x_3)|$ \rf{Edist}
is shown in Fig. 2 at $Q\bar Q$-separation 2 fm.
One can see at the figure the peaks of the color-Coulomb field \rf{3.10} over the quark and antiquark,
and the string \rf{3.13}  between them, with the universal profile ${\cal E}(\rho)$,
\be
{\cal E}(\rho)=2\sigma \lt(1+\fr{\rho}{\lambda}\rt)
\exp\lt(-\fr{\rho}{\lambda}\rt),
\lb{profE}
\ee
where $\rho$ is the distance to the $Q\bar Q$ axis.

\subsection{Magnetic currents distribution and Londons equation}

To perform more detailed analysis of the magnetic currents distribution
 \rf{k1} in the case of static quark and antiquark let us apply
the first Maxwell equation  \rf{Meq} to the electric field in bilocal
approximation \rf{Edist}, \rf{3.10}, \rf{3.13}. Then one
can seethat the magnetic current $\vek$ has a form
\be
\vek=\rot\, \vecE,
\label{k}
\ee
while the magnetic charge is absent.
It is clear that the perturbative color-Coulomb field \rf{V2} does not contribute to \rf{k}.
A nonperturbative field  \rf{3.13} is directed along the quark-antiquark axis,
therefore magnetic current winds around the axis. In particular case of the saturated string \rf{profE}
 the polar component of the magnetic current  $k_\varphi$ takes a form
\be
k_\varphi(\rho)=-\fr{2\sigma \rho}{\lambda^2}
\exp\lt(-\fr{\rho}{\lambda}\rt).
\lb{profk}
\ee
One can see that the value of the current rises linearly near the axis and falls exponentially at large distances from it.

The vector distribution of magnetic currents in the case
of $Q\bar Q$-separation $R=2$ fm is shown in Fig. 3.
This distribution resembles the one of the electric superconducting
currents around the Abrikosov string in superconductors
\cite{ano1}, and is another hint in favor of dual 
superconductivity mechanism of confinement \cite{ano2}.
An exponential behavior of current and field at large distances means that
the dual Londons equation
\be
\rot~\vek=\lambda^{-2}\vecE
\lb{Me1}
\ee
is satisfied.
Indeed, the only component of the polar vector
$k_\varphi$ \rf{profk} is directed along z axes and has a form
\be
(\rot~\vek)_z(\rho)=\fr{1}{\rho}
\fr{\partial\rho k_\varphi}{\partial\rho}=
\gamma(\rho)\,\lambda^{-2}{\cal E}(\rho),
\lb{Me2}
\ee
where the universal profile ${\cal E}(\rho)$ is defined in
\rf{profE}, and function
\be
\gamma(\rho)=\fr{-2+\rho/\lambda}{1+\rho/\lambda}
\lb{Me3}
\ee
rises monotonically from $-2$ and at $\rho\gg \lambda$ tends to unity as
$\gamma(\rho)\approx 1-3\lambda/\rho$.

One concludes that the confinement mechanism is related to
cyclic magnetic currents \rf{profk} squeezing the electric field
into the tube of string with the exponential fall off outside it,
and satisfying the dual Londons equation \rf{Me2}, \rf{Me3}.

\subsection{Vacuum polarization and screening of the coupling constant}
We turn now
to the second Maxwell equation for the static quark and antiquark, the
Gauss law
\be
\dv\,\vecE=\rho,
\lb{3.6.1}
\ee
where the field $\vecE$ \rf{Edist} is the sum
\be
\vecE=\vecE^{D_1,\mr{oge}}+\vecE^{D_1,\mr{np}}+\vecE^D,
\lb{3.6.2}
\ee
and $\vecE^{D_1,\mr{oge}}$, $\vecE^D$ are defined in \rf{3.8},
\rf{3.10},\rf{3.13}, while the nonperturbative field 
$\vecE^{D_1,\mr{np}}$ ensures the exponential fall-off of the formfactor
$D_1$ at large distances. In this section we introduce additional
assumption about charge distribution.
This assumption is confirmed {\it a posteriori} by the lattice
results in abelian projected gauge theory
(compare e.g. the distributions of effective field and its
nonperturbative part  along $Q\bar Q$ axis in Figs. 6,7 with
corresponding distributions in Fig. 21 from paper \cite{B}).
We assume that the nonperturbative
contributions to the charge density cancel,
\be
\dv\,\vecE^{D_1,\mr{np}}=-\dv\,\vecE^D,
\label{3.6.6}
\ee
so that the charge density has a form
\be
\rho=4\pi C_F \alpha_s\,(\delta(\ver)-
 \delta(\ver-\veR)).
\label{3.6.5}
\ee
Using explicit expression for the field $\vecE^D$  \rf{3.13}, we
find from \rf{3.6.6}  the ``screening'' charge density
$\tilde \rho(r)$,
\be
\dv\,\vecE^{D_1,\mr{np}}=\tilde \rho(r)-\tilde \rho(|\ver-\veR|),
\label{3.6.7}
\ee
\be
\tilde \rho (r)=-\fr{2\sigma}{\pi \lambda^2}\,r\,
K_1\lt(\fr{r}{\lambda}\rt).
\label{3.6.8}
\ee
Relying on \rf{3.6.7},  \rf{3.6.8}, one calculates  the field
$\vecE^{D_1,\mr{np}}$,
\be
\vecE^{D_1,\mr{np}}=\fr{\tilde Q(r)\,\ver}{r^3}-
\fr{{\tilde Q}(|\ver-\veR|)\,(\ver-\veR)}{|\ver-\veR|^3},
\label{3.6.9}
\ee
where $\tilde Q$ is the ``screening'' charge,
\be
\tilde Q(r)=
\fr{2\sigma \lambda^2}{\pi}
\int_0^{r/\lambda}x^3\,K_1(x)dx,
\label{3.6.11}
\ee
It is obvious that there is no field at large distances from quark
and antiquark due to confinement, and the full charge
 $Q(r)$ defined as
\be
Q(r)= C_F\alpha_s(r)-\tilde Q(r),
\label{3.26}
\ee
turns to zero. The condition $\lt.Q(r)\rt|_{r\to \infty}=0$ leads to the
relation \cite{rep3}
\be
C_F\alpha_s=3\sigma \lambda^2
\label{3.27a}
\ee
between the strong coupling at large distances and
parameters $\sigma, \lambda$ responsible for confinement.
The behavior of the charge $Q(r)$ at standard values $\sigma=0.18$ GeV$^2$,
$\lambda=0.2$ fm and constant value $\alpha_s=0.42$ calculated from \rf{3.27a}
is shown in Fig. 4. The mean radius of the screening according to the figure
is of the order of 0.5 fm.

 The behavior of the strong coupling taking into account the background
confining fields was studied in  \cite{BPT5,BPT6} in the framework of the
background perturbation theory. It was shown that the confining background
leads to the modification of the logarithmic running of coupling according to
$\alpha_s(q^2)\to \alpha_s(q^2+m_B^2)$, where the "background mass"
$m_B\approx 1$ GeV$\approx \lambda^{-1}$ is related to the
energy of the valence gluon
excitation and the gluon correlation length. One can see that  at large
distances $r\gg \lambda$ the background coupling tends to constant
("freezes"), while at small ones it turns to the running coupling of the
ordinary perturbation theory.
The relation between parameters now takes the form
\be
C_F\alpha_s(\lambda)=3\sigma \lambda^2,
\label{3.27}
\ee
where $\alpha_s(\lambda)$ is the freezing value of background coupling equal to
the ordinary running coupling at scale $\lambda$. This equation relates two
alternative scale parameters of quantum theory, $\Lambda_{\mr{QCD}}$ and
the string tension $\sigma$\footnote{It was shown in
\cite{alpha1} that the value $\Lambda_{\mr{QCD}}=241$ MeV computed in lattice
\cite{Capitan} in $\bar \mr{MS}$ regularization scheme with $n_f=0$ corresponds to the
freezing value  $\alpha_s(\lambda)=0.42$, the latter being in complete
agreement with \rf{3.27}.}.

In Fig. 5 the background running coupling is shown
by dotted curve. The behavior of the running charge  $Q^{\mr{run}}(r)$
 is shown by solid curve. As one can see from the figure, the effective charge has
 a maximum at $r\approx 0.3$ fm.

Using standard values $\sigma=0.18$ GeV$^2$,
$\lambda=0.2$ fm and constant value  $\alpha_s=0.42$ we plot the following
field distributions. In Fig. 6 the projections of fields $\vecE^D(0,0,x_3)$,
$\vecE(0,0,x_3)$  and  $\vecE^{D_1,\mr{oge}}(0,0,x_3)$
into the quark-antiquark axis are
shown. Note that  the fields  $\vecE^{D_1,\mr{np}}$ and
$\vecE^{D_1,\mr{oge}}$ cancel in the middle of
the string. In Fig. 7 the  projections of fields  $\vecE^D(0,0,x_3)$, 
$\vecE^D(0,0,x_3) +\vecE^{D_1,\mr{np}}(0,0,x_3)$ and  $\vecE(0,0,x_3)$ 
onto the quark-antiquark axis are plotted. In Fig. 8 the vector
distribution of the displacement field $\vecE(x_1,0,x_3)$ is shown 
demonstrating that the field is squeezed in tube with the width of the 
order of $\lambda$. In Fig. 9 the vector distribution of the solenoid 
field $\vecE^D(x_1,0,x_3) +\vecE^{D_1,\mr{np}}(x_1,0,x_3)$ is plotted.

It is convenient to define the isotropic dielectric function 
$\varepsilon(r)$,
\be
\varepsilon(r)=\fr{Q(r)}{C_F\alpha_s(r)}.
\lb{3.33}
\ee
One can calculate that at large distances $r\gg \lambda$ it is
exponentially small,
\be
\lt.\varepsilon(r)\rt|_{r\to\infty}=
\fr{\sqrt{\pi}}{2}\lt(\fr{r}{\lambda}\rt)^{5/2}
\exp\left(-\fr{r}{\lambda}\right),
\lb{3.35}
\ee
indicating the disappearence of the color-Coulomb field both inside and
outside the string.

\section{Hadrons with three static sources}
\subsection{Green functions and W-loops}
Physical hadrons are nonlocal extended objects, therefore to construct
their Green functions one should use nonlocal quark and gluon operators 
\be
q^i(x,Y)\equiv q^j(x) \Phi^i_j(x,Y),
\lb{4.1}
\ee 
\be
g_a(x,Y)\equiv g_b(x) \Phi_{ab} (x,Y),
\label{4.2}
\ee
as well as the local one
$G^j_i(x) \equiv  g_a(x) t^{(a)j}_i$.
Here and in what follows $i,j,...=1,2,3$ are color indexes in
fundamental representation, and $a,b,...=1,..,8$ in adjoint one;
$g_a$ denotes the valence gluon operator of background perturbation 
theory \cite{BPT5,BPT6}, and $G^j_i(x)$ transforms as
$G_i^j\to U^{+j}_{{}j'}G_{i'}^{j'}U^{i'}_{i}$
under gauge transformations.
One can construct gauge invariant combinations of these operators using
symmetric tensors $\delta_i^j,~ \delta^{ab},~ d^{abc}$ and
antisymmetric ones $e_{ijk}$,  $f^{abc}$,
\be
B_Y(x,y,z,Y) =
e_{ijk}q^i(x,Y)
q^j(y,Y)q^k(z,Y),
\label{4.4}
\ee
\be
 G^{(f)}_Y(x,y,z,Y)=f^{abc}g_a(x,Y) g_b(y,Y) g_c(z,Y),
\label{4.5}
 \ee
\be
 G^{(d)}_Y(x,y,z,Y)=d^{abc}g_a(x,Y) g_b(y,Y) g_c(z,Y),
\label{4.53}
 \ee
\be
G_\Delta(x,y,z) =G_i^j (x) \Phi_j^k(x,y)
G_k^l(y) \Phi_l^m (y,z) G_m
^n(z) \Phi_n^i(z,x).
\label{4.6}
\ee
First three constructions have a structure of $Y$-type with the
string junction at point $Y$, where the color indexes are contracted 
with the (anti-) symmetric tensor, and the latter one has
a structure of triangular type. Let us stress that the wave function of 
triangular type is possible only for glueballs but not
for baryons, see \cite{hadrons}.

Hadron Green function has a form
\be
{\cal G}_i(\bar X, X)=\lan \Psi^+ _i(\bar X)\Psi_i(X)\rrr,
\label{4.7}
\ee
where $\Psi_i=G_\Delta, G_Y, B_Y;~X=x,y,z$ in the case of $G_\Delta$ and
 $x,y,z,Y$ for $Y$-states.
 The vacuum average $\lan...\rrr$ leads
 to the product of Green functions  of quarks or valence gluons,
 which are proportional to the parallel transporters,
$$
\lan \bar q_j (\bar x) q^i(x)\rrr\sim
\Phi_j^i (\bar x, x),
$$
\be
 \lan  g_a(\bar x) g_b(x)\rrr\sim \Phi_{ab}(\bar x, x).
\label{4.8}
\ee
Therefore the hadron Green function is proportional to the W-loop of
this hadron, see equation \rf{ev157}, and is reduced to the W-loop in
the case of static sources. W-loops of baryon and $Y$-type glueball
take forms correspondingly
\be
{\cal W}_B=\frac16 \lan  \epsilon_{ijk}
\epsilon^{i 'j 'k '}
\Phi^{i}_{i '}(C_1)\Phi^{j}_{j '}(C_2)
\Phi^{k}_{k '}(C_3)\rrr,
\lb{4.9}
\ee
\be
{\cal W}_G^{Y,f}=\frac1{24}\lan f^{abc} f^{a'b'c'}
\Phi^{aa'}(C_1)\Phi^{bb'}(C_2)
\Phi^{cc'}(C_3)\rrr,
\lb{4.10}
\ee
\be
{\cal W}_G^{Y,d}=\frac3{40}\lan d^{abc} d^{a'b'c'}
\Phi^{aa'}(C_1)\Phi^{bb'}(C_2)
\Phi^{cc'}(C_3)\rrr.
\lb{4.11}
\ee
Trajectories $C_i$ formed by the sources are shown in Fig. 10.
A W-loop of $\Delta$-type glueball at large distances 
can be represented as a product of three meson W-loops \cite{hadrons}, 
\be
{\cal W}_G^\Delta(X,\bar X)=W(\bar x,\bar y|x,y)
W(\bar y,\bar z|y,z) W(\bar z,\bar x|z,x).
\lb{4.12}
\ee
Corresponding contours are shown in Fig. 11.

\subsection{Static potentials}
Static potentials of hadrons with three static sources  are calculated
 in bilocal approximation of the field correlator method
 \cite{hadrons,Vpot}
   in the same way as meson ones\footnote{The effect of charge screening
is taken into account in \cite{talk}}. For hadrons of $Y$-type let us denote
$\ven^{(a)}$ the unit vector directed from the string junction to the
$a$-th quark and $R_a$ the separation between this quark and the string
junction. The potential in baryon reads
\be
V_B(R_1,R_2,R_3)=
\left(\sum_{a=b}-\sum_{a<b}\right)n_i^{(a)}n_j^{(b)}\int_0^{R_a}\int_0
^{R_b } d l\,d l' \int_0^\infty  d t\, {\cal D}_{i4,j4}(z_{ab}),
\lb{4.13}
\ee
where $z_{ab}=(l\,\ven^{(a)}-l'\ven^{(b)},t)$.
One can represent this potential in the form
\be
V_B=V^{c}+V^d+V^{\mr{nd}},
\lb{4.13a}
\ee
where $V^{c}$ is the color-Coulomb potential
\be
V^{c}= -\fr{C_F \alpha_s}2 \sum_{i<j}\fr1{r_{ij}},
\lb{4.13b}
\ee
$r_{ij}$ is the $i$-th and $j$-th quark separation.
We will take into account the charge screening replacing
in \rf{4.13b} $C_F \alpha_s$ with $Q$ defined in
 \rf{3.26}, \rf{3.6.11}. Terms $V^d$ and $V^{\mr{nd}}$ denote
 the diagonal and nondiagonal parts of the potential
 corresponding to the correlator $D$; $V^d$ is determined
 by the first and  $V^{\mr{nd}}$ by the second sum in \rf{4.13}.
One can find explicit expressions for  $V^d$ and $V^{\mr{nd}}$
in \cite{Vpot}. We just note here that  $V^d$ is a sum of
quark-antiquark potentials $V^D$ (\ref{ev960}), \rf{ev36},
\be
V^d(R_1,R_2,R_3)=\sum_a V^D(R_a).
\lb{4.13c}
\ee
Characteristic feature of the potential \rf{4.13} is an increase of
its slope when the source separations are increasing. In Fig. 12
the behavior of the baryon potential with the color-Coulomb part 
subtracted is shown in comparison with the lattice data \cite{Tak} as a 
function of the total length of baryon string $L=\sum_a R_a$. A tangent 
with the slope $\sigma$ is shown by points. One can see from the figure 
that the potential slope becomes significantly less than $\sigma$ at 
$L\la 1$ fm. This effect is induced by the influence of the correlation 
length of confining fields \cite{Vpot}. In Fig. 13 a dependence of the
baryon potential in equilateral triangle on the quark separation is given 
in comparison with the lattice data \cite{deF}. Note the agreement
between analytic and lattice calculations within the accuracy of a few 
tens MeV.
For the $Y$-glueball potential the Casimir scaling holds,
\be
\fr{V_G^Y}{V_B}=\fr{C_8}{C_3},
\lb{4.14}
\ee
where  $C_3=(N_c^2-1)/2N_c\equiv C_F$ and $C_8=N_c$ are quadratic
Casimir operators in fundamental and adjoint representations.

A potential in  $\Delta$-glueball in the case of equilateral triangle
with the side $r$ has a form \cite{hadrons}
\be
V_G^\Delta(r)=\fr{C_8}{C_3}V^{c}(r)+V^d(r)-
2V^{\mr{nd}}(r).
\lb{4.15}
\ee
Let us note that $V^d$ and $V^{\mr{nd}}$  depend on the valence gluons
separation but not on the separation between the gluon and the center of 
the triangle, and that the term $-2V^{\mr{nd}}$ corresponds to the 
interaction of three effective quark-antiquark W-loops.
The behavior of the potentials $V_G^Y$ and $V_G^\Delta$
in equilateral triangle in dependence on the source  separation
$r$ is shown in Fig. 14. The potential  $V_G^Y$ goes above
the   $V_G^\Delta$ because of the positive contribution of nondiagonal 
term $V^{\mr{nd}}$ to $Y$-type glueball and negative to triangular one, 
as well as for the greater slope of diagonal term $V^d$ in the case of
$Y$-glueball.

\subsection{Fields distributions}
  The field in baryon is defined  \cite{rep3} as the square average
 \be
(\vecE^{(B)})^2=\frac23\, \lt((\vecE^B_{(1)})^2
+(\vecE^B_{(2)})^2+(\vecE^B_{(3)})^2\rt)
\lb{4.16}
\ee
of fields $\vecE^B_{(i)}$  calculated for the probe plaquette
joint to the trajectory $C_i$,
\be
\vecE^B_{(1)}(\vex,\veR^{(1)},\veR^{(2)},\veR^{(3)})=
\vecE^M(\vex,\veR^{(1)})-\frac12\,
\vecE^M(\vex,\veR^{(2)})-\frac12\,\vecE^M(\vex,\veR^{(3)}).
\lb{4.17}
\ee
Normalizing coefficient 2/3 in \rf{4.16} is chosen due to the
condition that at large separations the field acting on quarks equals
to $\sigma$. According to \rf{4.16}, \rf{4.17}, the field in baryon
 is expressed through fields of effective quark-antiquark pairs, with
 positions of antiquarks coinciding with the string junction.
The distribution of the field  $\vecE^{(B)}$ taking into account
only the contribution of formfactor $D$ is shown in Figs. 15 and 16
in the plane of quarks forming an equilateral triangle with the side
1 fm and 3.5 fm respectively. One can see in Fig. 16 three
plateau with the saturated profile, and small growth of the field
around the string junction point, the relative difference of values
amounts to 1/16. A surface formed by the confining field with the value
$\sigma$ is shown in Fig. 17 for quark separations 1 fm. One can see 
the small convexity in the region of the string junction.

A field in  $\Delta$-type glueball is a sum of meson fields with
gluon pairs acting as the effective sources \cite{rep3},
\be
\vecE_\Delta^{(G)}(\vex,\ver^{(1)},\ver^{(2)},\ver^{(3
)})=
\sum_{i=1}^3 \vecE^M(\vex-\ver^{(i)},
\ver^{(i+1)\mr{mod3}}-\ver^{(i)}),
\lb{4.18}
\ee
where $\ver^{(i)}$ denotes the position of $i$-th valence gluon.
In Fig. 18 the field distribution $|\vecE_\Delta^{(G)}(\vex)|$
in valence gluon plane is shown at gluon separations 1 fm, and
in Fig. 19 the surface $|\vecE_\Delta^{(G)}(\vex)|=\sigma$ is plotted 
for the same gluon separations. Let us note that according to
 \rf{3.13b} one can calculate the static quark-antiquark potential
 as the work of the force acting on the quark, done on separating the 
latter from the antiquark to the distance $R$. Analogous relation are 
valid for the field and potential of 
 $\Delta$-type glueball, see  \rf{4.15}, \rf{4.18}.
It is clear that the nondiagonal part of the $\Delta$-glueball potential
 $V^{\mr{nd}}$ equals to the work of force acting on the effective quark
 from the external string and is therefore related to the interference
 of the meson fields  $\vecE^M$ in the vicinity
 of valence gluons  of the order of $\lambda$.

\section{Conclusions}

In this paper we have systematically treated: the vacuum fields in QCD,
the confinement mechanism, the QCD string formation and finally, the
field distribution inside hadrons. Everywhere we have used the field
correlators as a universal gauge-invariant formalism, which allows to 
describe all phenomena appearing in QCD. In description of vacuum fields
the most important property is the Gaussian dominance: the lowest 
(Gaussian) correlator is dominating on the minimal area surface of the
W-loop, and there are sufficient grounds for the statement that the
total distribution of higher correlators does not exceeed few percent.
This phenomenon, found on the lattice \cite{bali21}, is not yet fully
understood, see \cite{i9,i13}, although it gives an explicit 
dynamical picture, which possibly is incompatible with the old physics
of the instanton gas, of $Z_2$-fluxes $etc.$. Therefore, one can assert
that the picture of the maximally stochastic QCD vacuum is a very good
approximation to the reality. One can remember that the measure of 
coherence is associated with the weight of the contribution of higher
correlators, e.g. for the instanton gas the total contribution of higher 
(non-Gaussian) correlators is dominating. Moreover, the vacuum
correlation length $\lambda$ (i.e. the factor in the exponent for the
asymptotics of the Gaussian correlator) is relatively small, 
$\lambda\sim 0.2$  fm for the quenched vacuum. This value is much
smaller than the typical hadron radius, $\sim 1$ fm. Theoretically, the
smallness of $\lambda$ is connected to a large mass gap for glueballs
and gluelumps, since  $\lambda=1/M$, where $M$ is the lowest gluelump
mass, $M\sim 1.4$ GeV, which is calculated both analytically and on the 
lattice \cite{i111,i112,i113}.

 Let us turn now to the confinement
mechanism. From the point of view of field correlators, confinement
occurs due to the appearence of a specific term in the Gaussian 
correlator, denoted $D(x^2)$, which violates Bianchi identities in the
abelian case and therefore is absent in case of QED. If, however, one
considers the $U(1)$ theory with magnetic monopoles present in the 
vacuum, then the function $D(x^2)$ is nonzero and is proportional to the
monopole current correlator. The next step is to find the source of
$D(x^2)$ (i.e. the source of confinement) in the nonabelian theory. It
was done in \cite{uy1,uy2}, where the derivatives of $D(x^2)$ were
connected to the triple correlator $\lan EEB\rrr$. Thus the problem of 
establishing of the confinement mechanism in the formalism of field
correlators reduces to the problem of calculating   $D(x^2)$ and the 
triple correlator and to the finding the conditions of 
its appearence/disappearence (e.g. in QCD -- as functions of temperature 
or baryon density). The lattice calculations confirm the disappearence
of $D(x^2)$ at the deconfinement temperature $T_c$, and with it have
confirmed all cofinement picture in the framework of the present method.
One expects that at the next step-- by computing correlators (including  
$D(x^2)$) with the help of the gluelump Green's functions in the whole
$x$ region -- one will make the field correlator method selfconsistent,
and the problem of confinement will be solved quantitatively and in 
principle. 

At the same time, this universal formalism of field 
correlators can be used to study the distribution of effective fields
and currents, defined with the help of the W-loop. This representation,
see chapter 3, enables one to describe, on one hand, the dual Meissner
effect \cite{i151,i152}, and on the other hand, it relates to the
effective Lagrangian approach of Adler and Piran \cite{ap} and
dielectric vacuum models, \cite{lee51,lee52} and subsequent papers.
Indeed, the field correlator method not only admits this approximate 
qualitative interpretation, but also yields explicit expressions for the
density of effective electric charges and effective magnetic currents.
Being the gradient of the color-Coulomb potential at small distances,
the effective field condenses into a tube on the characteristic hadron
scale and ensures confinement. In the process the strong coupling
constant is screened due to the vacuum polarization by nonabelian 
gluonic interactions.

Finally, let us summarize the contents of the last
chapter devoted to field distributions inside hadrons with three 
constituents (sources). Here the field correlator method is the only
quantitative analytic method, and its comparison with numerical
(lattice) results is very interesting. One can note that in the method 
one has only two parameters -- the string tension $\sigma$ and the 
correlation length $\lambda$, $\lambda$ being expressed through
$\sigma$, and   $\sigma$ is playing the role of the scale parameter
related to $\Lambda_{\mr{QCD}}$.  The baryon potential computed in this
way \cite{Vpot} is in good agreement with lattice calculations and gives
an independent confirmation that baryon strings have the structure of
the $Y$-type with the string junction. Moreover, the field correlator 
method explains the smaller slope of the baryon potential at the typical
hadron distances, known from the baryon phenomenology -- the decrease of 
the slope is caused by the string interference effects connected to
nonzero correlation length  $\lambda$. The three-gluon glueballs, in
contrast to baryons, can have the structure of both $Y$-type and
$\Delta$-type \cite{hadrons}. However, the latter is preferred
energetically. In the concluding part of the last chapter the field
distributions in baryons and in the  $\Delta$-type glueballs are given,
where one can visualize the shape of the string in these hadrons. 

Summarizing, one can say that the universal language of the field
correlator method turns out to be extremely convenient in all cases
considered. In particular, it enables one to formulate the
gauge-invariant description of the QCD vacuum as some medium with
properties ensuring confinement. 

The authors are grateful to L.B.Okun 
for his support, to N.O.Agasian, M.I.Polikarpov for
discussions and to A. Di Giacomo, V.G.Bornyakov and D.V.Antonov for
useful remarks. This work was supported by the grant INTAS 00-110. D.K.
and Yu.S.  are also supported by the grant  INTAS 00-00366. V.Sh. is
grateful to FOM and Dutch National Scientific Fund (NOW) for a financial
support.

\newpage

\begin{figure}[!t]
\hspace*{1.5cm}
\epsfxsize=10cm
\epsfbox{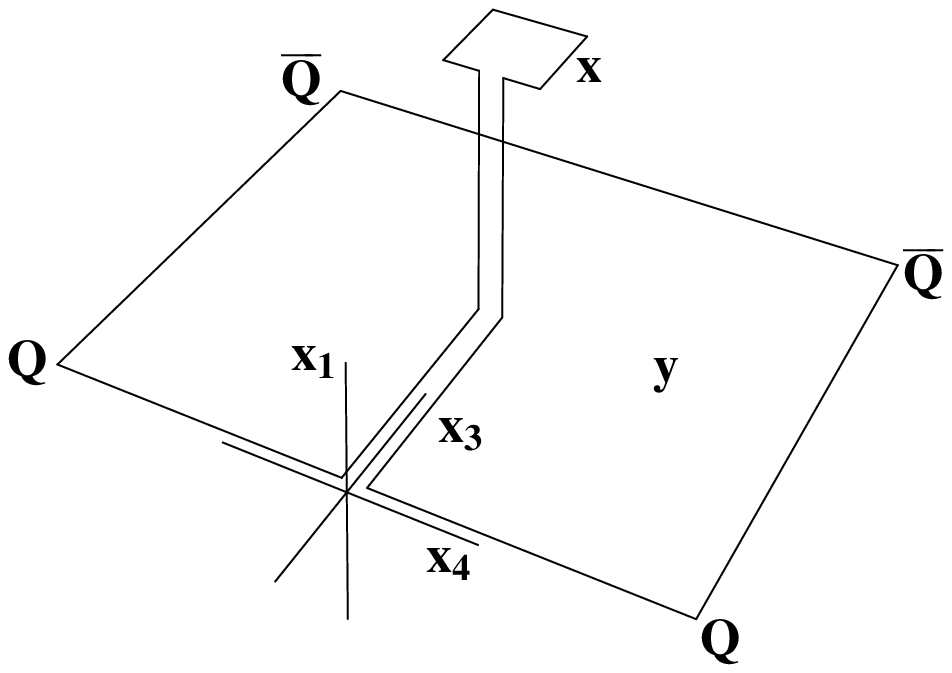}
\caption{ A connected probe \rf{F2} for static quark and antiquark}
\end{figure}

\begin{figure}[!b]
\hspace*{-1cm}
\epsfxsize=16cm
\epsfbox{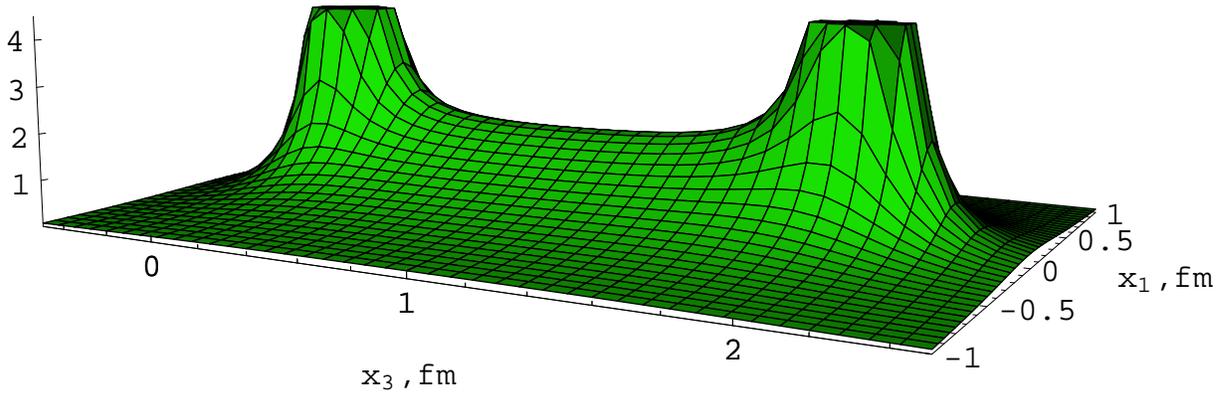}
\caption{ A distribution of the field  
 $|\vecE(x_1,0,x_3)|$ \rf{Edist} at quark-antiquark separation 2 fm.
Cutted peaks of color-Coulomb field and string between 
quark and antiquark are clearly distinguished. The standard 
values of parameters $\sigma=0.18$ GeV$^2$,
$\lambda=0.2$ fm are used.}
\end{figure}

\clearpage

\begin{figure}[!t]
\hspace*{3cm}
\epsfxsize=7cm
\epsfbox{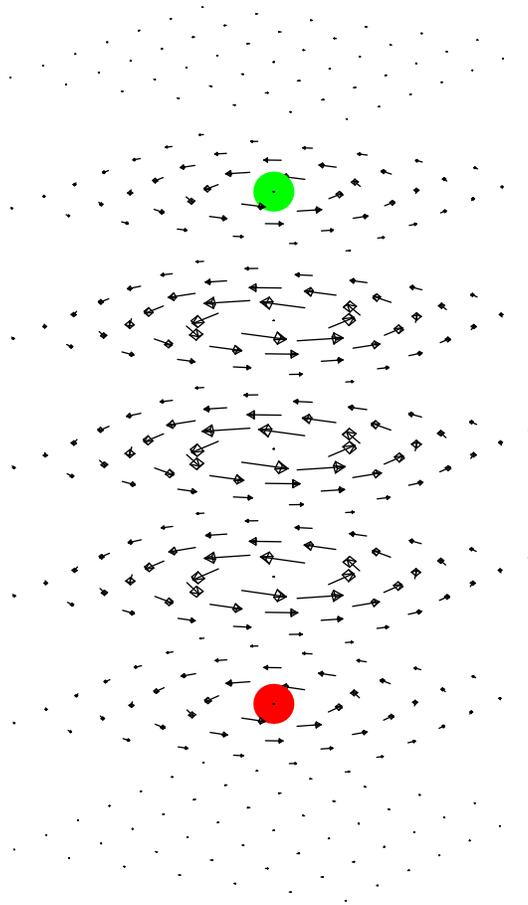}
\caption{ A vector distribution of magnetic currents \rf{3.13}, 
\rf{k} at quark-antiquark separation 2 fm.
Positions of quark and antiquark are shown by points. }
\end{figure}

\clearpage

\begin{figure}[!t]
\epsfxsize=10cm
\epsfbox{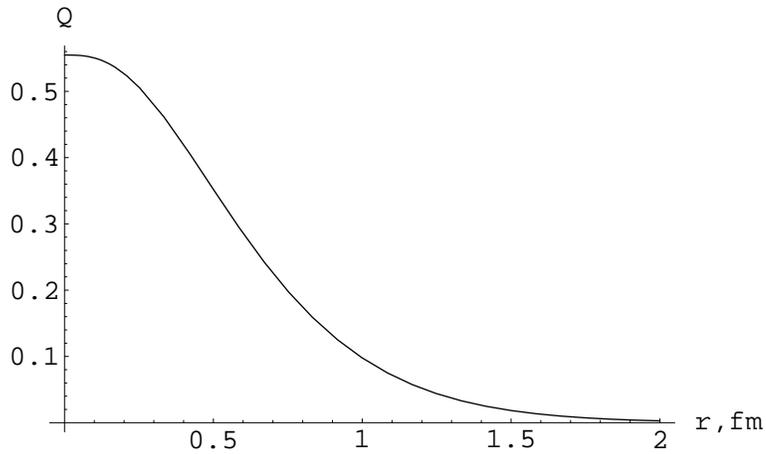}
\caption{ An effective charge $Q(r)$ \rf{3.26} in dependence of
the distance from the quark for $\sigma=0.18$ GeV$^2$,
$\lambda=0.2$ fm  and constant value $\alpha_s=0.42$. }
\end{figure}

\begin{figure}[!t]
\epsfxsize=10cm
\epsfbox{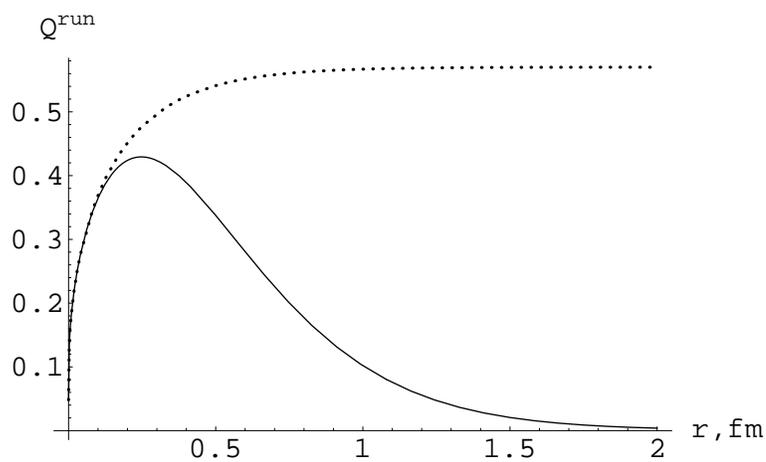}
\caption{ A running background coupling $C_F \alpha_B(r)$
\cite{alpha1} (dotted curve) and running effective charge
$Q^{\mr{run}}=C_F \alpha_B(r)-\tilde Q(r)$ (solid curve) vs. the 
distancefrom the quark.}
\end{figure}
\clearpage

\begin{figure}[!t]
\hspace*{0.5cm}
\epsfxsize=12cm
\epsfbox{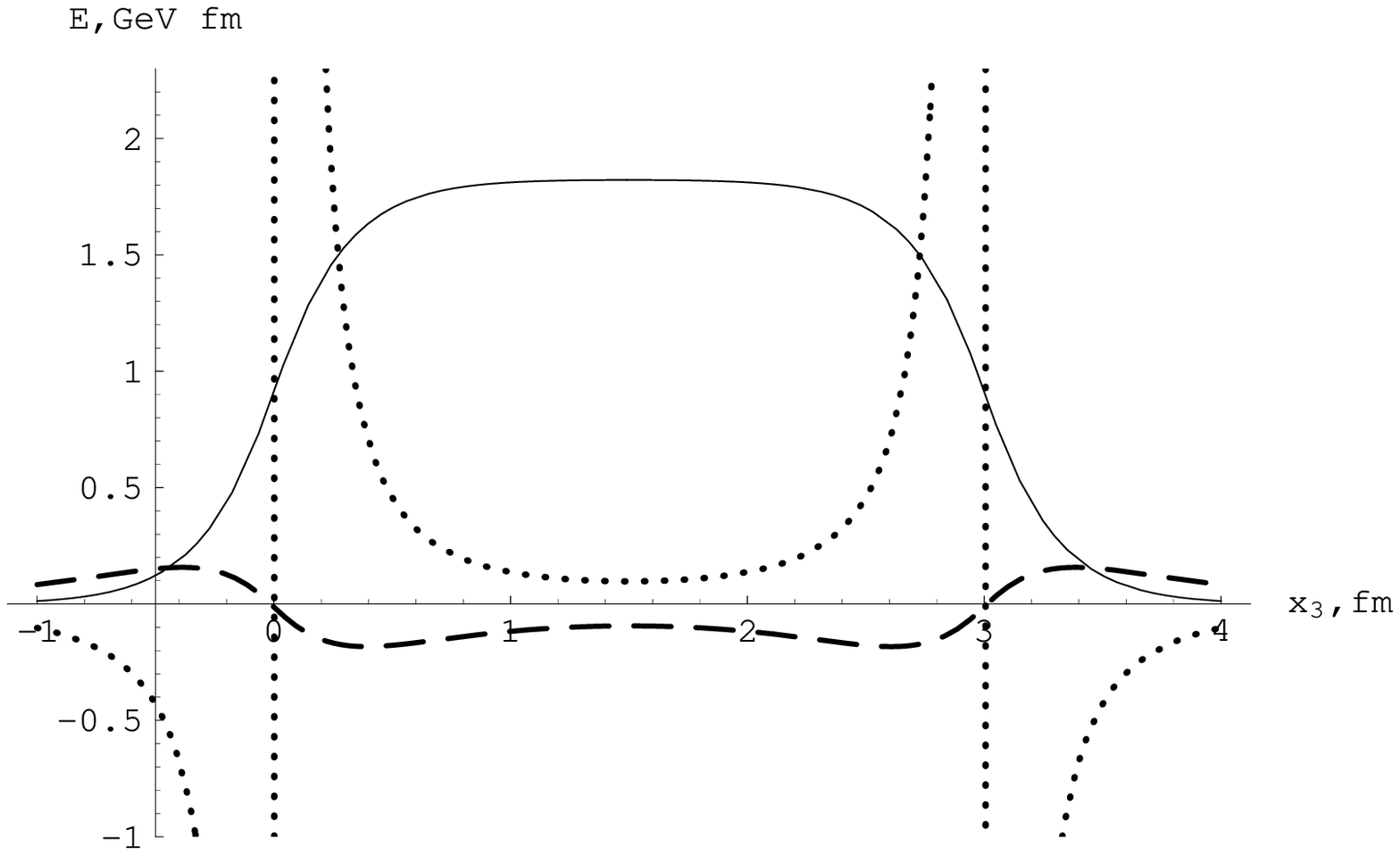}
\caption{ Distributions of the projections of the fields 
 $\vecE^D(0,0,x_3)$  (solid curve),  $\vecE^{D_1,\mr{np}}(0,0,x_3)$ 
(dashed curve) and  $\vecE^{D_1,\mr{oge}}(0,0,x_3)$ (dotted curve) onto 
the quark-antiquark axis at $Q\bar Q$-separation 3 fm.}
\end{figure}

\begin{figure}[!t]
\hspace*{0.5cm}
\epsfxsize=12cm
\epsfbox{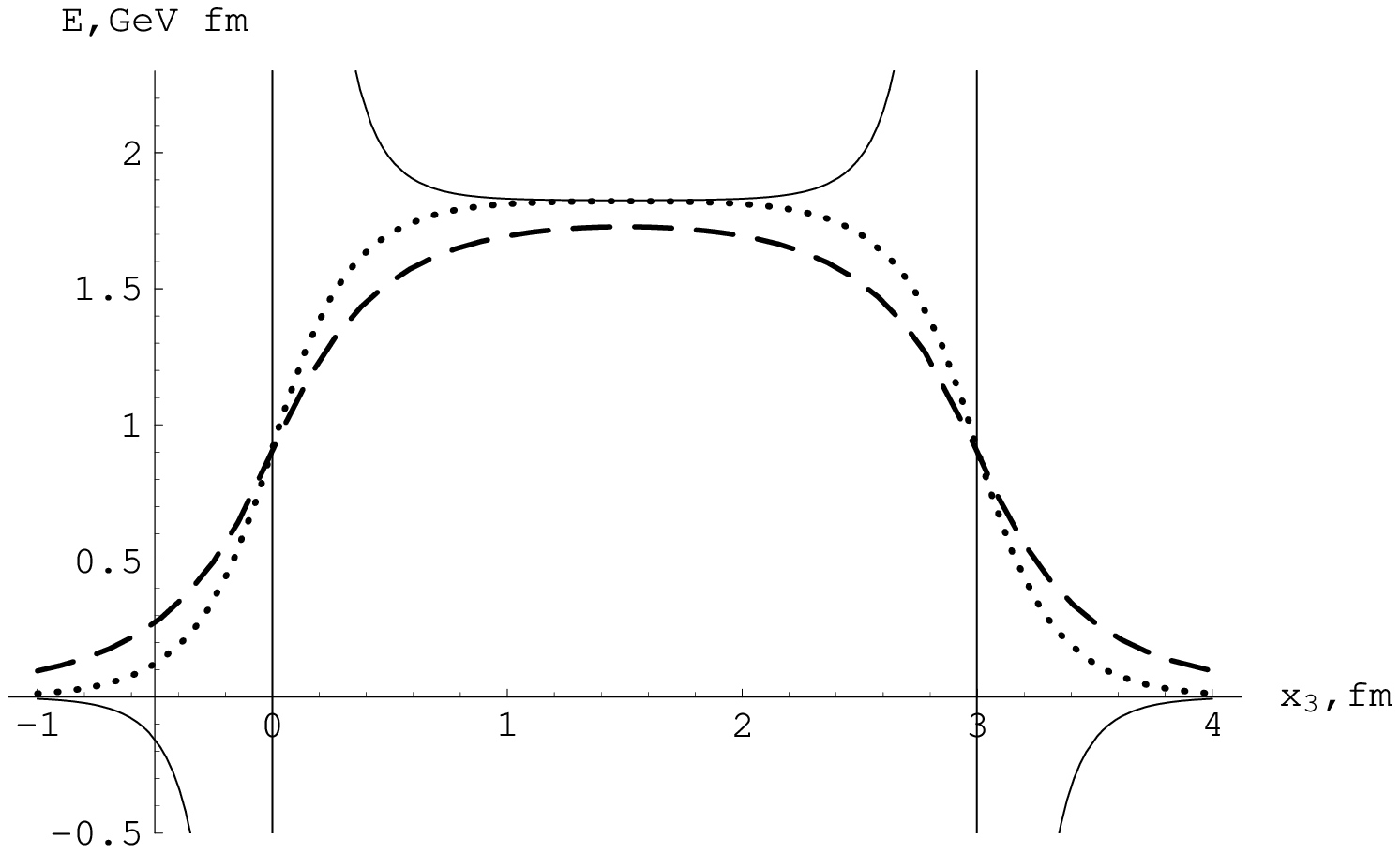}
\caption{ Distributions of the projections of the fields  
$\vecE^D(0,0,x_3)$ (solid curve), $\vecE^D(0,0,x_3) 
+\vecE^{D_1,\mr{np}}(0,0,x_3)$ (dashed curve)  and  $\vecE(0,0,x_3)$ 
(dotted curve)  into the quark-antiquark axis at $Q\bar Q$-separation 3 
fm.}\end{figure}

\clearpage

\begin{figure}[!t]
\hspace*{-2cm}
\epsfxsize=16cm
\epsfbox{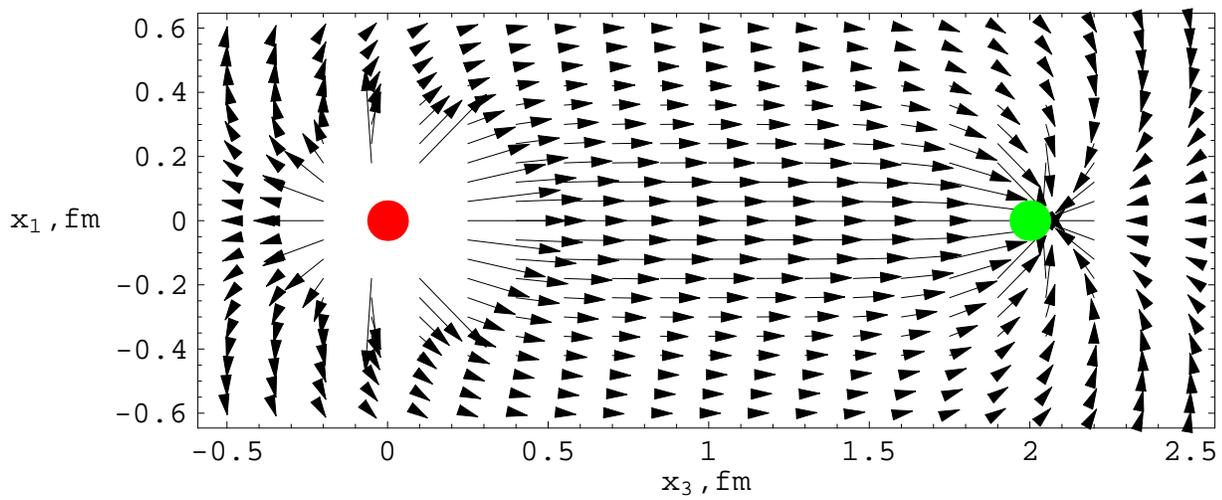}
\caption{ Vector distribution of the field
$\vecE(x_1,0,x_3)$. Positions of quark and antiquark are marked by 
points. }\end{figure}

\begin{figure}[!t]
\epsfxsize=12cm
\epsfbox{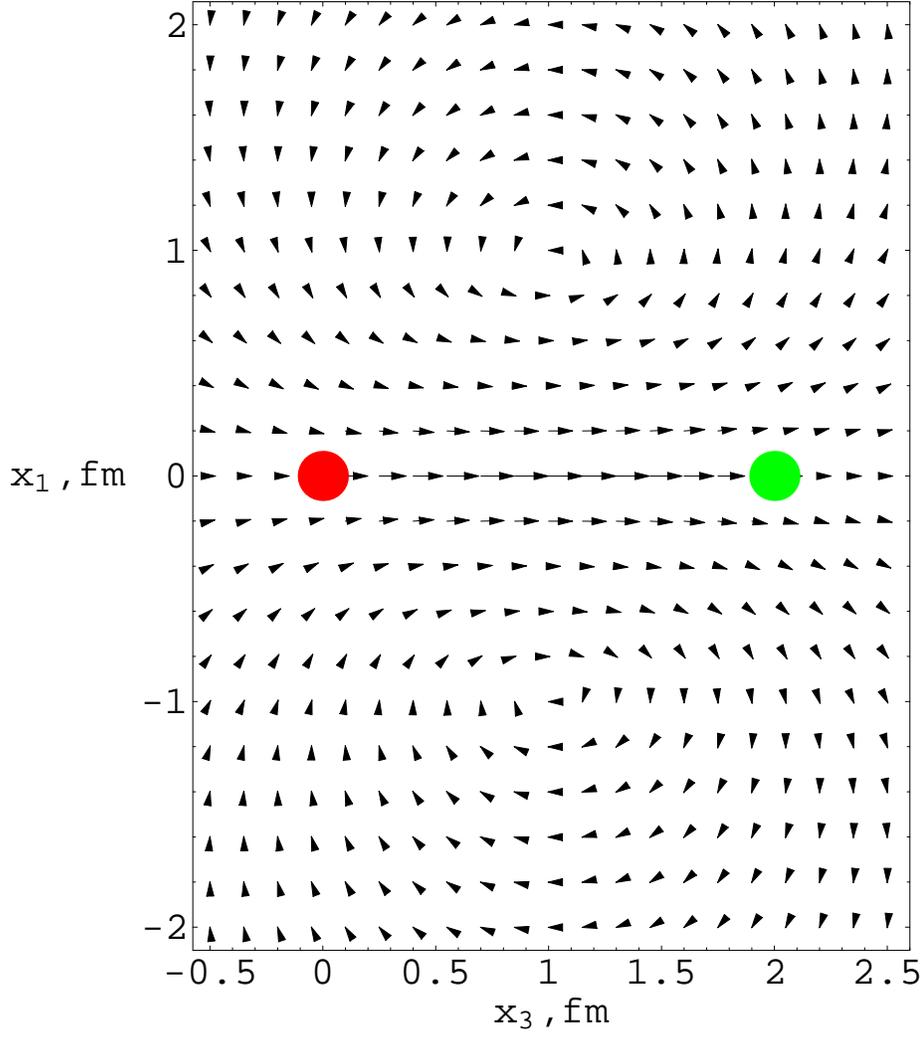}
\caption{Vector distribution of the solenoid field   $\vecE^D(x_1,0,x_3) 
+\vecE^{D_1,\mr{np}}(x_1,0,x_3)$. Positions of quark and antiquark are 
marked by points.}\end{figure}

\clearpage

\begin{figure}[!t]
\hspace*{1.5cm}
\epsfxsize=10cm
\epsfbox{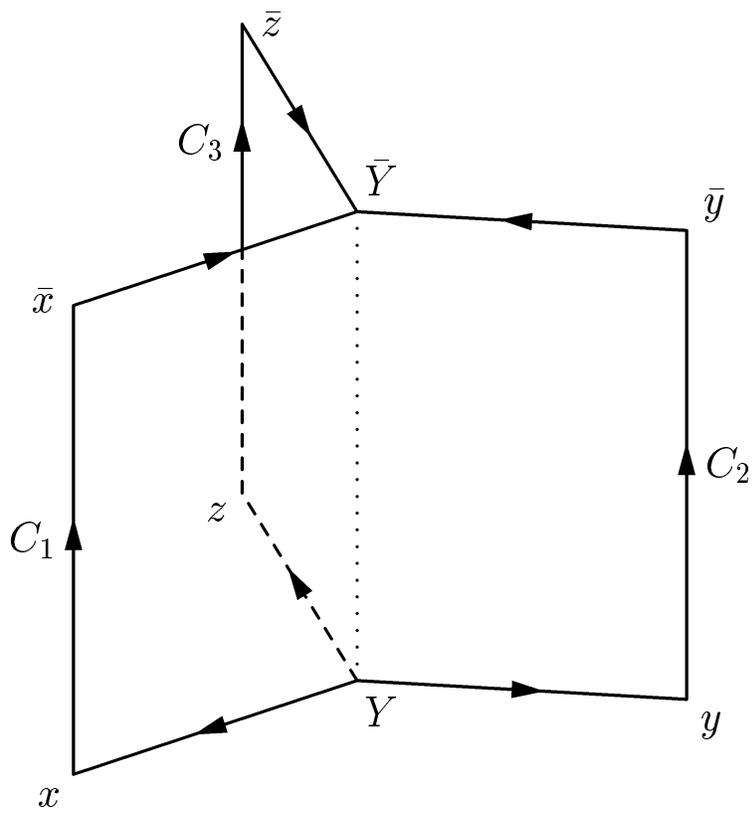}
\caption{ A W-loop of $Y$-type.}
\end{figure}

\begin{figure}[!t]
\hspace*{1.5cm}
\epsfxsize=10cm
\epsfbox{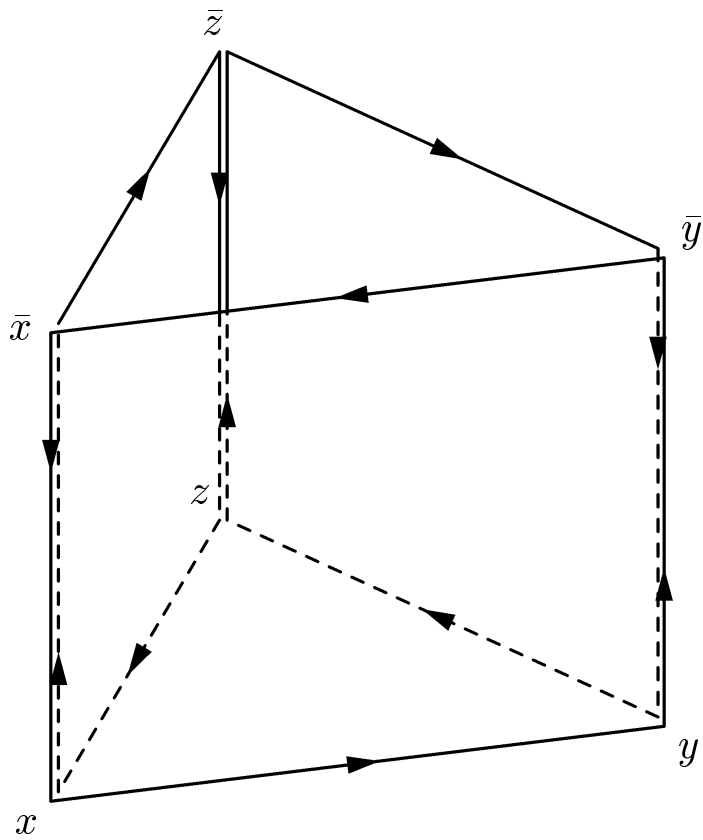}
\caption{ A W-loop of $\Delta$-type.}
\end{figure}

\clearpage

\begin{figure}[!t]
\epsfxsize=12cm
\epsfbox{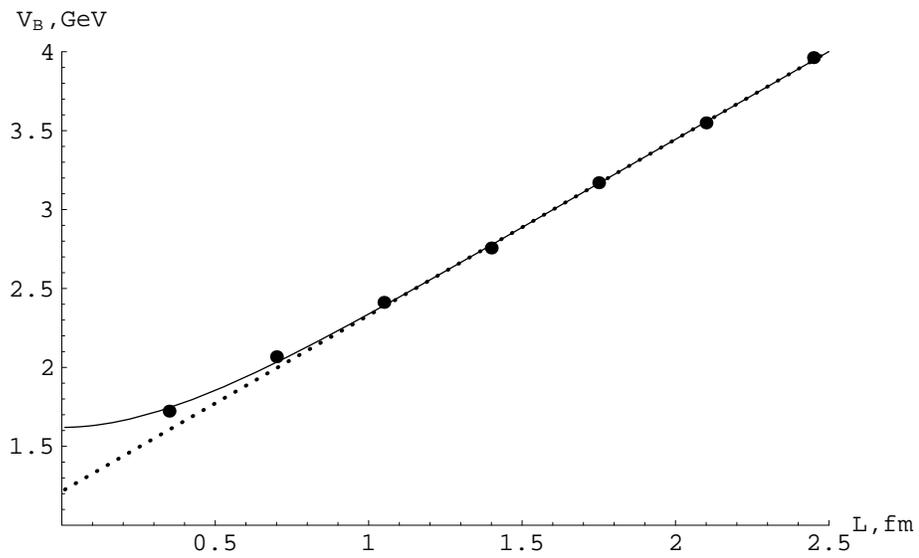}
\caption{A potential in baryon \rf{4.13} with the color-Coulomb part
contracted (solid curve) in comparison with the lattice data
  \cite{Tak} points in dependence on the total length of the baryon string
 $L$. A value of the string tension is $\sigma=0.22$ GeV$^2$. According to
\rf{3.27}, the corresponding value of correlation length is $\lambda=0.18$ fm.}
\end{figure}

\begin{figure}[!t]
\epsfxsize=12cm
\epsfbox{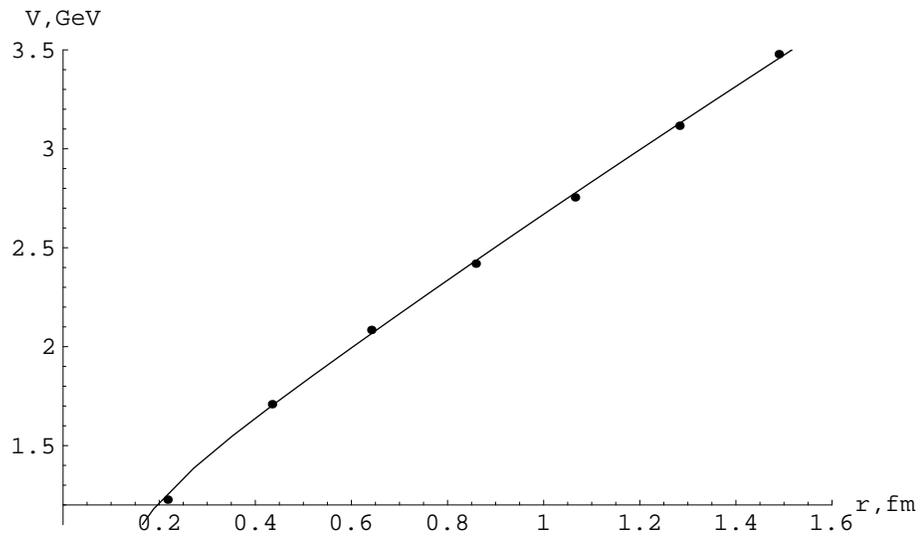}
\caption{ A dependence of the baryon potential in equilateral triangle
on quark separation $r$ (solid curve) in comparison with  the lattice data
 \cite{deF} (points).  A value of the string tension is
$\sigma=0.17$ GeV$^2$ ($\lambda=0.21$ fm).}
\end{figure}

\clearpage

\begin{figure}[!t]
\epsfxsize=12cm
\epsfbox{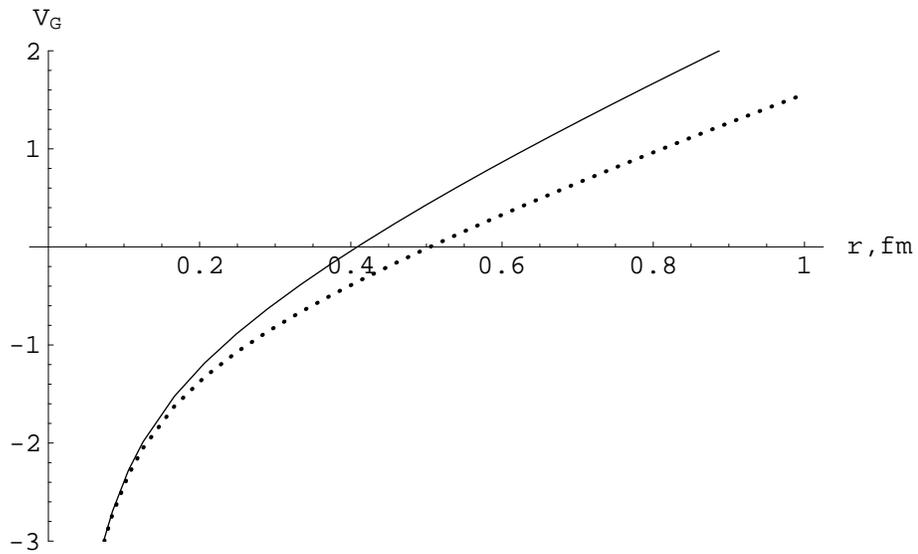}
\caption{ Potentials of three-gluon-glueballs
 $V_G^Y$ (solid curve) and $V_G^\Delta$ (dotte curve)
 in equilateral triangle vs. the sources separation $r$.}
\end{figure}

\clearpage

\begin{figure}[!t]
\hspace*{-0.5cm}
\epsfxsize=14cm
\epsfbox{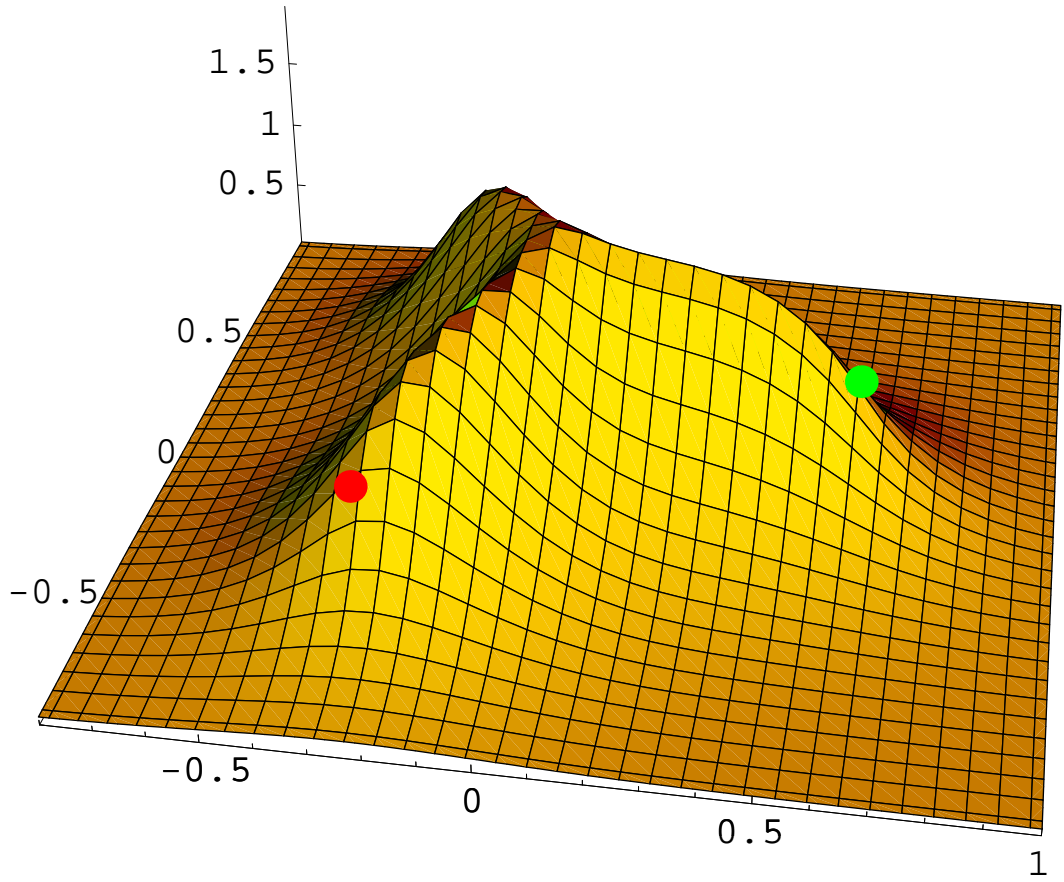}
\caption{A distribution of the field $\vecE^{(B)}$ \rf{4.16}, \rf{4.17} 
in GeV/fm with the only correlator  $D$ contribution considered in the
quark plane for equilateral triangle with the side 1 fm. 
Coordinates are given in fm, positions of quarks are marked
by points. }
\end{figure}

\begin{figure}[!t]
\hspace*{-0.5cm}
\epsfxsize=14cm
\epsfbox{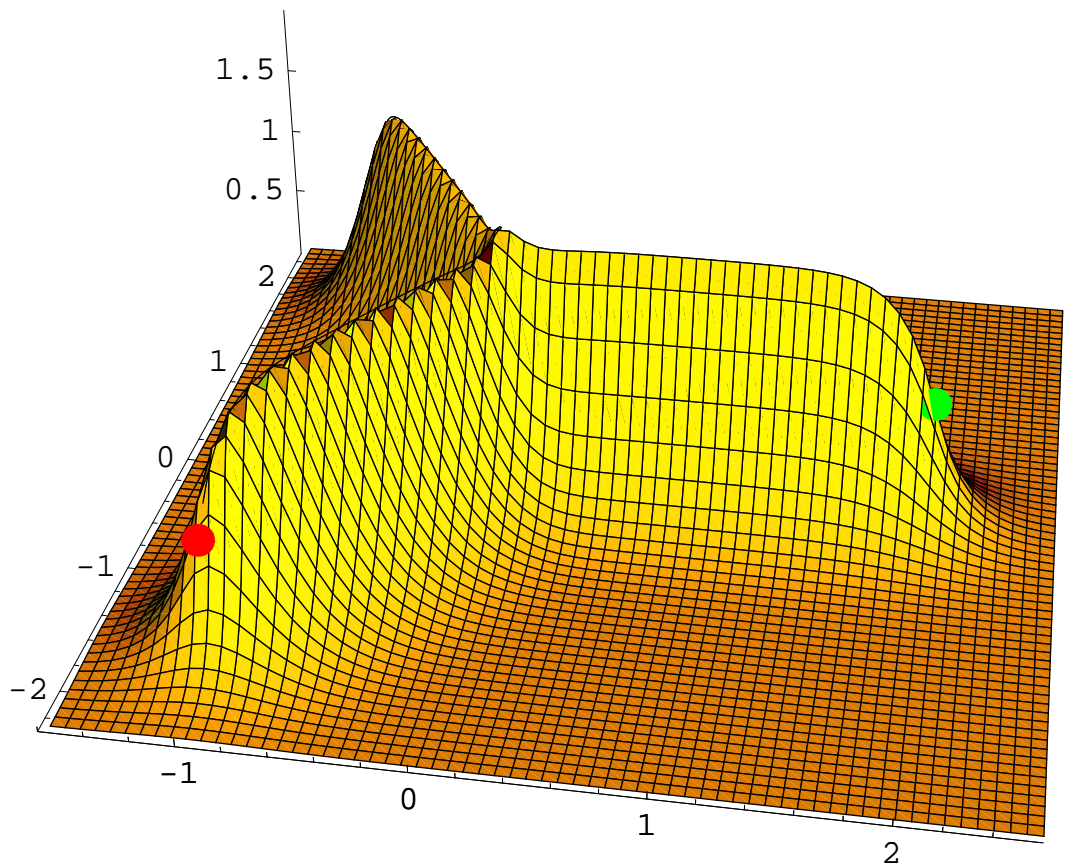}
\caption{ The same as in Fig. 15 but with the side of the
equilateral triangle equal to 3.5 fm. }
\end{figure}

\clearpage

\begin{figure}[!t]
\hspace*{-1cm}
\epsfxsize=15cm
\epsfbox{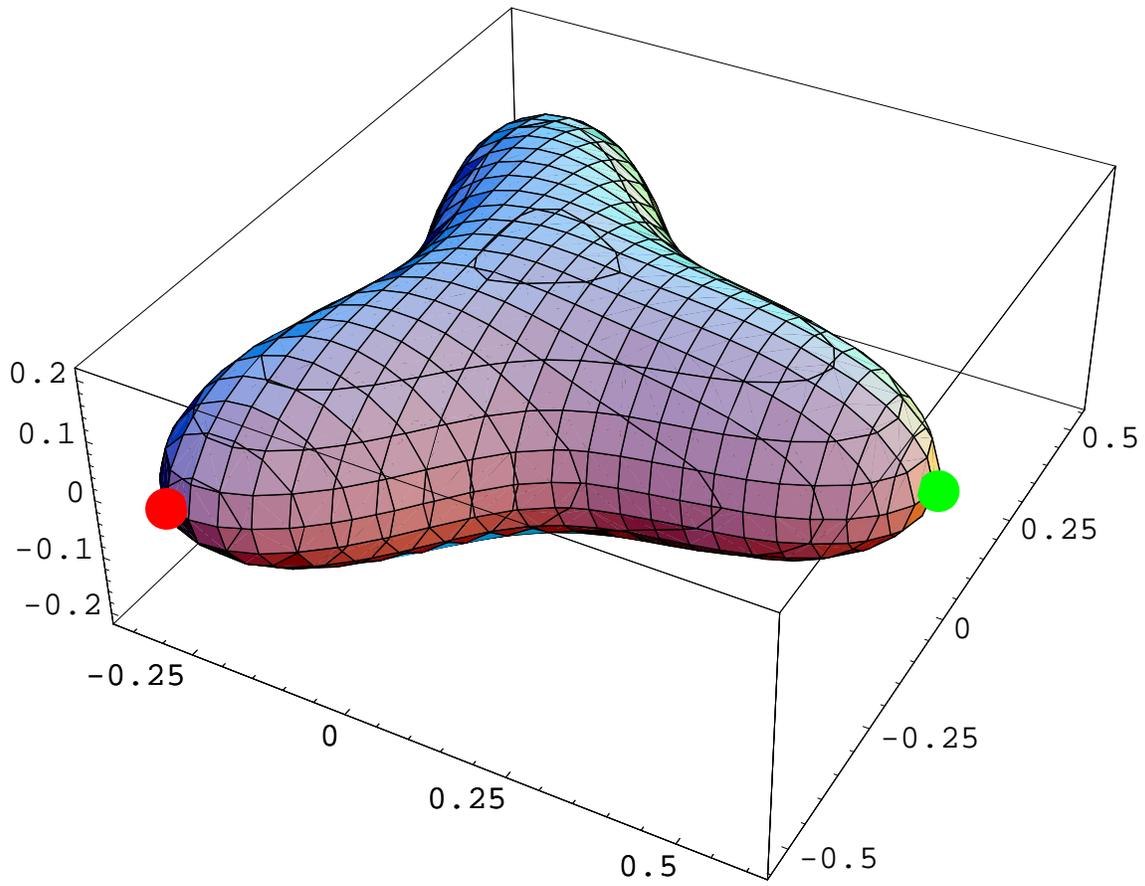}
\caption{ A surface $|\vecE^{(B)}(\vex)|=\sigma$ at quark separations
1 fm. Coordinates are given in fm, positions of quarks are marked
by points. }
\end{figure}

\clearpage

\begin{figure}[!t]
\hspace*{-0.5cm}
\epsfxsize=14cm
\epsfbox{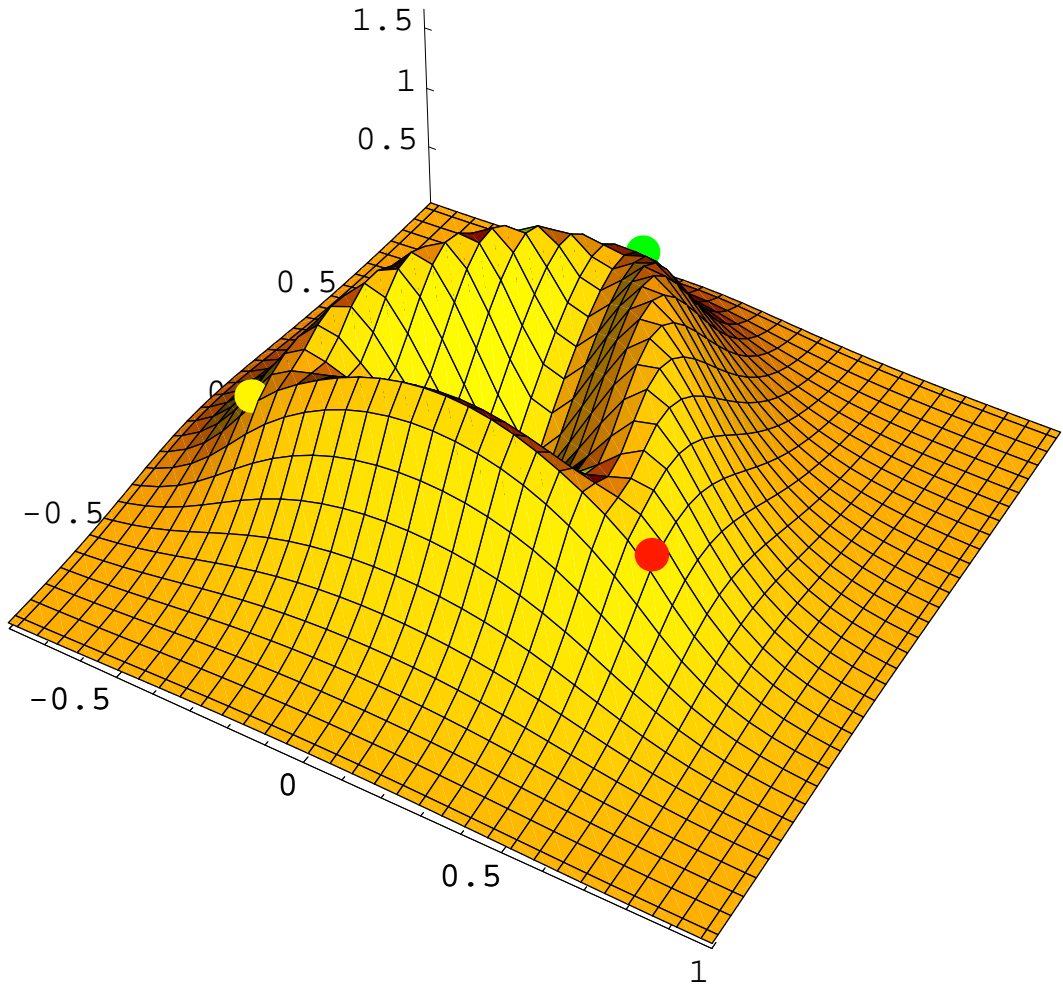}
\caption{ A distribution of the field   $|\vecE_\Delta^{(G)}(\vex)|$ 
\rf{4.18} in GeV/fm of the triangular glueball in the plane of valence gluons
with separations 1 fm. Coordinates are given in fm, positions of valence gluons are marked
by points. }
\end{figure}

\begin{figure}[!t]
\hspace*{-1cm}
\epsfxsize=15cm
\epsfbox{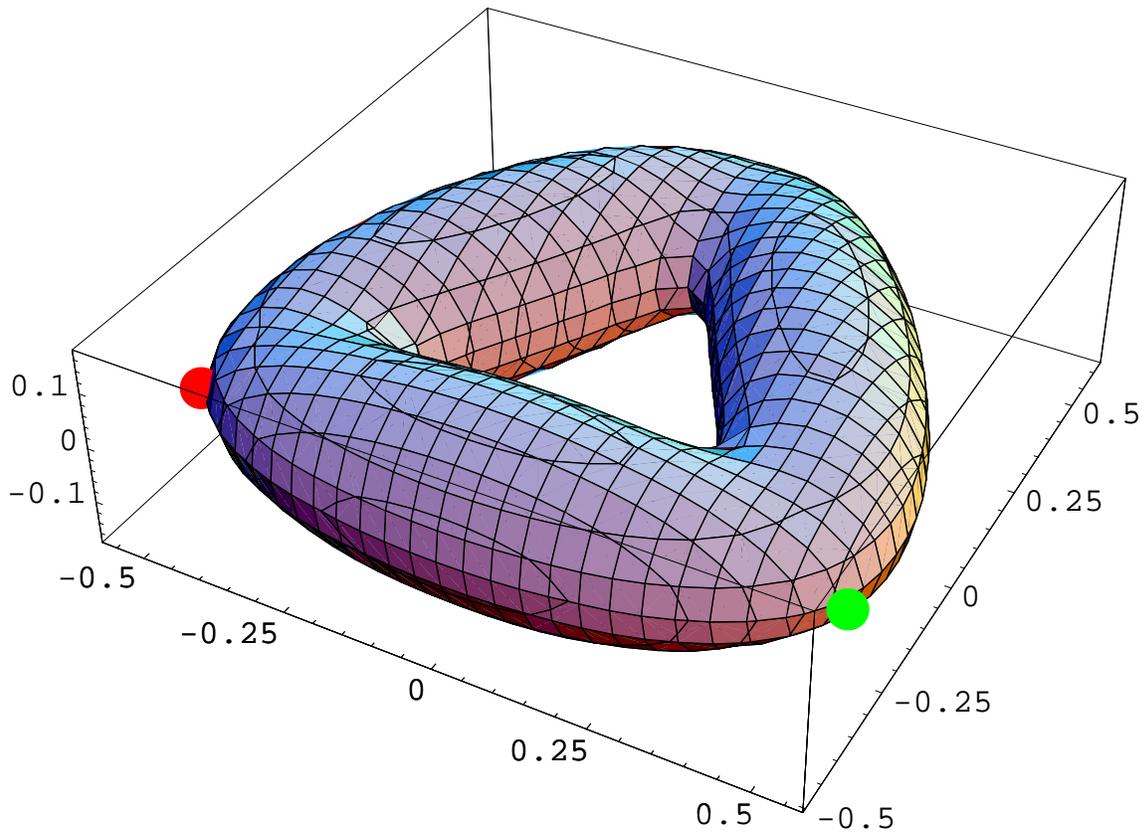}
\caption{ A surface $|\vecE_\Delta^{(G)}(\vex)|=\sigma$
 at valence gluons separations 1 fm. Coordinates are given in fm,
 positions of valence gluons are marked by points.}
\end{figure}

\end{document}